\documentclass[twocolumn]{aastex701}
\graphicspath{{./}{figures/}}

\begin{document}

\title{7DT Insight: Variability in Young Stellar Objects}

\author[0000-0002-1408-7747]{Mi-Ryang Kim} 
\affiliation{Astronomy Program, Department of Physics and Astronomy, Seoul National University, Seoul 08826, Republic of Korea}
\email{mi.ryang@snu.ac.kr} 

\author[0000-0003-3119-2087]{Jeong-Eun Lee}
\affiliation{Astronomy Program, Department of Physics and Astronomy, Seoul National University, Seoul 08826, Republic of Korea}
\affiliation{SNU Astronomy Research Center, Seoul National University, Seoul 08826, Republic of Korea}
\email[show]{lee.jeongeun@snu.ac.kr} 

\author[0000-0002-8537-6714]{Myungshin Im}
\affiliation{Astronomy Program, Department of Physics and Astronomy, Seoul National University, Seoul 08826, Republic of Korea}
\affiliation{SNU Astronomy Research Center, Seoul National University, Seoul 08826, Republic of Korea}
\email{myungshin.im@gmail.com}

\author[0000-0003-4010-6611]{Jinho Lee}
\affiliation{Department of Electrical and Computer Engineering, Seoul National University, Seoul 08826, Republic of Korea}
\email{leejinho@snu.ac.kr}

\author[0000-0002-1418-3309]{Ji Hoon Kim}
\affiliation{Astronomy Program, Department of Physics and Astronomy, Seoul National University, Seoul 08826, Republic of Korea}
\affiliation{SNU Astronomy Research Center, Seoul National University, Seoul 08826, Republic of Korea}
\email{jhkim.astrosnu@gmail.com}

\author[0000-0002-3118-8275]{Seo-Won Chang}
\affiliation{Astronomy Program, Department of Physics and Astronomy, Seoul National University, Seoul 08826, Republic of Korea}
\affiliation{SNU Astronomy Research Center, Seoul National University, Seoul 08826, Republic of Korea}
\email{seowon.chang@gmail.com}

\author[0000-0002-6639-6533]{Gregory S.~H. Paek}
\affiliation{Institute for Astronomy, University of Hawaii, 2680 Woodlawn Drive, Honolulu, HI 96822, USA}
\affiliation{Astronomy Program, Department of Physics and Astronomy, Seoul National University, Seoul 08826, Republic of Korea}
\affiliation{SNU Astronomy Research Center, Seoul National University, Seoul 08826, Republic of Korea}
\email{gregorypaek94@gmail.com} 

\author[0000-0003-4422-6426]{Hyeonho Choi}
\affiliation{Astronomy Program, Department of Physics and Astronomy, Seoul National University, Seoul 08826, Republic of Korea}
\affiliation{SNU Astronomy Research Center, Seoul National University, Seoul 08826, Republic of Korea}
\email{hhchoi1022@gmail.com} 

\author[0000-0002-9852-2469]{Donggeun Tak}
\affiliation{Astronomy Program, Department of Physics and Astronomy, Seoul National University, Seoul 08826, Republic of Korea}
\affiliation{SNU Astronomy Research Center, Seoul National University, Seoul 08826, Republic of Korea}
\email{takdg123@gmail.com}

\author[0000-0002-0786-7307]{Donghwan Hyun}
\affiliation{Astronomy Program, Department of Physics and Astronomy, Seoul National University, Seoul 08826, Republic of Korea}
\affiliation{SNU Astronomy Research Center, Seoul National University, Seoul 08826, Republic of Korea}
\email{hdhd333@gmail.com}

\author[0009-0005-6140-8303]{Won-Hyeong Lee}
\affiliation{Astronomy Program, Department of Physics and Astronomy, Seoul National University, Seoul 08826, Republic of Korea}
\affiliation{SNU Astronomy Research Center, Seoul National University, Seoul 08826, Republic of Korea}
\email{wohy1220@gmail.com}

\author[0000-0003-3130-7921]{Hyeyoon Lee}
\affiliation{Department of Electrical and Computer Engineering, Seoul National University, Seoul 08826, Republic of Korea}
\email{hylee817@snu.ac.kr}

\author[0009-0002-1079-8178]{ShinGeon Kim}
\affiliation{Department of Electrical and Computer Engineering, Seoul National University, Seoul 08826, Republic of Korea}
\email{happyriosshs@snu.ac.kr}

\author[0000-0001-7629-3573]{S. Thomas Megeath}
\affiliation{Ritter Astrophysical Research Center, Dept. of Physics and Astronomy, University of Toledo, Toledo, OH 43606, USA}
\email{s.megeath@utoledo.edu} 

\begin{abstract}
Photometric variability in young stellar objects (YSOs) provides critical insight into the mechanisms of mass accretion, disk evolution, and circumstellar extinction in early stellar evolution. We present an analysis of day-timescale optical variability in the Orion A central region using two-night 7-Dimensional Telescope (7DT) medium-band photometry obtained on March 23 and 24, 2024. The 7DT observations provide optical spectral sampling with 16 medium-band filters spanning 400--825 nm, enabling direct two-epoch comparisons. To remove satellite-trail contamination, we used an SSIM-based ResNet classifier (accuracy 0.97; F1 = 0.93) to exclude affected exposures. Subsequent photometry and two-epoch variability measurements yielded a working sample of 769 YSO candidates, among which we identified 110 variables ($\sim$14\%), including seven extreme cases with $|\Delta m_\lambda|>0.5$ mag. To describe the wavelength dependence of the variability, we compared five simple templates: extinction-like changes ($R_V =$ 3.1 and 5.5), a gray (wavelength-independent) change, and two spot-like toy models (hot and cold) implemented as two-temperature surface mixtures. 
The best-fit results are dominated by spot-like templates (37 cold-spot and 22 hot-spot objects), with 37 sources best matched by extinction-like templates and 14 by the gray template. The m650 excess fraction is higher in the hot-spot and gray templates than in the others. This could be compatible with more frequent line/veiling-related contributions in those groups, although the m650 excess is not a direct accretion diagnostic.
\end{abstract}


\section{introduction}
T Tauri stars are low-mass young stellar objects (YSOs) that provide important probes into early stellar and planetary formation \citep{1945Joy}. These pre-main-sequence (PMS) stars are more luminous than their main-sequence counterparts because they are still undergoing gravitational contraction. Some, particularly classical T Tauri stars, also exhibit ongoing accretion from circumstellar disks. These disks supply material for accretion and can drive winds or jets. 

T Tauri stars are generally divided into classical T Tauri stars (CTTSs), which are often identified as Class II sources and show strong emission lines and infrared excesses due to active accretion, and weak-line T Tauri stars (WTTSs), which correspond to the Class III stage and typically exhibit weaker emission lines and lack significant infrared excess, consistent with more evolved or dissipated inner disks \citep{1988Walter,1989Bertout,2016Hartmann}.

Photometric variability is a defining characteristic of T Tauri stars, spanning timescales from minutes to decades. Variations on rotational timescales are typically associated with stellar spots rotating in and out of view, while more irregular or short-term changes can arise from variable accretion, disk occultation, or magnetic flaring events \citep{1992Attridge,1996Choi,2007Bouvier,2013Romanova,2013Bouvier,2014Stauffer,2023Bino}.
In CTTSs, this variability is often linked to changes in the structure of the inner disk and to fluctuations in accretion activity, as supported by statistical correlations between variability properties and UV excess, H$\alpha$ emission, and other disk diagnostics identified in previous monitoring studies \citep{2014Venuti,2015Venuti,2014Stauffer,2017Rigon}.
More broadly, accretion in YSOs varies over a continuous range of amplitudes and timescales, from low-level, short-term fluctuations to large-amplitude, long-duration outbursts. This behavior is now recognized as a fundamental aspect of how stars accumulate their final mass \citep{2023Fischer}.

A widely accepted framework for explaining the accretion process in CTTSs is the magnetospheric accretion model. In this scenario, strong stellar magnetic fields disrupt the inner disk and funnel disk material along magnetic field lines toward the stellar surface \citep{1991Koenigl}. The infalling material accelerates under gravity and forms accretion columns, which terminate in shocks when they reach the stellar surface. These shocks generate compact hot regions, often called hotspots, that produce excess continuum emission most prominently in the ultraviolet and blue-optical wavelength range \citep{1998Calvet,2016Hartmann}. The energetic processes associated with magnetospheric accretion contribute not only to photometric variability but also to complex spectroscopic signatures, including broad permitted emission lines and, in some cases, redshifted absorption components seen in H I, He I, Ca II, and other accretion-sensitive transitions (e.g. \citealt{1994Hartmann,1998Muzerolle,2001Muzerolle,2013Salyk,2014Alcala,2017Alcala,2015Rigliaco,2020Komarova,2024Rogers,2025Tofflemire,2025Fiorellino}).

The configuration of the stellar magnetic field also plays a significant role in shaping the variability. In many CTTSs, the magnetic axis is tilted relative to the rotation axis, leading to non-axisymmetric accretion flows and localized hotspots. This geometry concentrates the accretion at particular locations and can result in rotational modulation of hotspot visibility \citep{2012Alencar,2021Espaillat}. It may also influence the detectability of inverse P Cygni profiles and other line-profile diagnostics, as well as the overall variability behavior.

Optical time-series observations remain a fundamental tool for investigating T Tauri stars. Emission lines, such as H$\alpha$, H$\beta$, the Ca II triplet, and several He I lines (e.g., 587.6 nm, 667.8 nm), serve as tracers of accretion and outflow activity \citep{2006Natta,2014Alcala,2017Alcala,2024Rogers,2025Fiorellino}, while time-series photometry provides a powerful diagnostic of surface inhomogeneities, variable extinction, and changes in accretion behavior \citep{2001Carpenter,2016Bozhinova,2019Akimoto,2021Froebrich}. Medium-band optical photometry, in particular, yields improved sensitivity to continuum slopes and flux variations in line-sensitive bands compared to broad-band filters, making it an effective method for disentangling different variability mechanisms in young stars.

Time-domain studies of YSOs have relied on two complementary approaches, each with distinct limitations. 
Multi-epoch spectroscopy, including targeted X-shooter monitoring of individual systems, provides simultaneous diagnostics of accretion, extinction, and line-profile variability across a broad wavelength range (e.g., \citealt{2018Schneider}). Beyond such single-target studies, larger X-shooter-based efforts such as PENELLOPE have also been used to investigate accretion-related variability in young stars (e.g., \citealt{2021Manara,2022Claes,2023Armeni}). In parallel, near-infrared surveys such as APOGEE/IN-SYNC have obtained multi-epoch H-band spectra for large young stellar samples, including the Orion population analyzed by \citet{2016DaRio,2017DaRio}, providing repeated spectroscopic data that are also well suited for variability-oriented analyses, even when variability was not the primary survey design. 

Broad-band photometric monitoring campaigns, including earlier large-sample UV/optical studies of young stars in NGC 2264 and more recent multi-band campaigns, enable statistical studies of larger samples and periodicity analysis \citep{2014Venuti,2015Venuti,2024Wendeborn}.
However, broad-band filters integrate flux over wide wavelength ranges, often blending continuum variations with multiple accretion-sensitive optical lines, including H I, He I, Ca II, Na I, and O I transitions \citep{2014Alcala,2017Alcala,2025Fiorellino}. This low spectral resolution leads to degeneracies where different physical mechanisms (e.g., variable extinction or changing accretion rates) can produce similar color slopes in color-magnitude diagrams (CMDs), necessitating reliance on assumed extinction laws or additional spectroscopic data to break the degeneracy \citep{2014Cody,2015Venuti,2015McGinnis,2024Wendeborn}. 

The 7-Dimensional Telescope (7DT; \citealt{2024Im}) bridges the gap by providing low-resolution spectroscopic imaging through a dense set of medium-band filters. Utilizing 20 filters spaced by 25 nm in the current configuration, 7DT enables low-resolution (R $\sim$ 20--40) spectral energy distribution (SED) sampling for a large number of sources across a wide field of view of 1.3 deg$^2$. This capability allows simultaneous low-resolution SED sampling for large numbers of sources across the field, enabling characterization of wavelength-dependent variability ($\Delta m_\lambda$) for substantially larger samples than is typically feasible with slit spectroscopy. We note that the working sample analyzed in this paper is defined by our detection and selection criteria.

In this study, we investigate the day-timescale variability of YSO candidates in the Orion A central region using optical imaging data obtained with the 7DT telescope. We construct a photometric catalog based on observations in 16 medium-band filters, cross-match the sources with existing YSO catalogs, and examine their wavelength-dependent variability properties.

\section{the Orion A central region}
The observed field covers the central region of the Orion A molecular cloud, including Sh 2-279, OMC 1--4, and the Orion Nebula. This area is structured by the Integral-Shaped Filament (ISF; \citealt{1987Bally}), a prominent filamentary feature extending over $\sim$10 pc. 

Located at a distance of $d \approx 414 \pm 7$ pc \citep{2017Kounkel}, the ISF represents one of the nearest active sites of clustered star formation. It contains a rich population of YSOs spanning a wide range of evolutionary stages, and its dense backbone is traditionally divided into the Orion Molecular Clouds (OMC) 1 through 4.
Millimeter and submillimeter observations have revealed a continuous dust ridge extending along these cores \citep{1997Chini}. 

In the center part of the filament lies OMC 1, the most active site of high-mass star formation in Orion A, which harbors the Orion Nebula Cluster (ONC) and the deeply embedded Kleinmann-Low (KL) nebula. Extending northward, OMC 2 and OMC 3 are associated primarily with low- and intermediate-mass star formation but show different source demographics. OMC 3 contains a larger fraction of deeply embedded Class 0/I protostars and cold condensations \citep{2000Aso}, whereas OMC 2 contains a relatively larger population of more evolved, disk-bearing Class II sources. This contrast has often been interpreted as an evolutionary gradient along the filament.

The field also includes the Sh 2-279 complex, encompassing NGC 1973, NGC 1975, and NGC 1977. \citet{2008Peterson} describe NGC 1977 as part of a continuous star-forming complex with OMC-2/3 and the Orion Nebula, while \citet{2024Boyden} show that it hosts several hundred pre-main-sequence stars in an intermediately irradiated environment dominated by B-type stars rather than O-type stars. The close juxtaposition of embedded filamentary populations and a less embedded nebular cluster further highlights the environmental diversity of the central Orion A cloud.

This combination of structural continuity and environmental diversity makes the Orion A central region well suited for studies of pre-main-sequence variability. Infrared surveys have established a broad census of YSOs across the region \citep{2012Megeath}, with Class II sources being of particular interest for this study. Characterized by optically thick disks and moderate accretion, these objects represent an important stage of disk evolution. In regions of moderate extinction, many of these systems are accessible to deep optical observations. This enables precise photometric and variability studies that probe stellar rotation, accretion, and circumstellar obscuration while strongly complementing existing infrared diagnostics across different local environments.

\section{Observation and data reduction}
\subsection{7DT Observations}
The observations were conducted using the 7-Dimensional Telescope (7DT), which is remotely operated by Center for the Gravitational-Wave Universe at Seoul National University \citep{2024SPIE13099E..1ZK, 2024SPIE13094E..0XK,2024ChoiHH}. Located at the El Sauce Observatory in Chile, 7DT currently consists of sixteen 50-cm modified Cassegrain telescopes. Each unit is equipped with a CMOS camera (9576 $\times$ 6388 pixels), providing a field of view (FoV) of 1.33$^\circ$ $\times$ 0.89$^\circ$. The main scientific goal of 7DT is to carry out follow-up observations for transient objects, particularly the electromagnetic counterparts of gravitational-wave events. This is enabled by its wide FoV and low-resolution spectroscopic imaging capability. While the full system is designed to employ 40 medium-band filters spanning 400--900 nm in 12.5 nm intervals, the currently operational 16 telescopes utilize a suite of 20 filters with 25 nm intervals. Filters are designated by the prefix "m" followed by their central wavelength in nanometers (for example, m400 corresponds to a filter center at 400 nm).

Observations of the Orion A central region were carried out over two nights on March 23 and 24, 2024. Individual exposures were taken with an integration time of 60 s per frame, and the total exposure times quoted below represent the cumulative integration of these frames. In this observation, sixteen filters from m400 to m825 were observed for a total exposure time of 2 hr per filter, split into 1 hr per night, while the m850 and m875 filters were observed for 1 hr during a single night. Due to an instrument configuration error, the m550 and m575 filters were not observed. Because our study focuses on overnight variability, we excluded the data from the m850 and m875 filters, which were observed on only a single night. Thus, the variability investigated in this paper corresponds to short-term, day-scale variability. 

\subsection{Data reduction}
The data stream from 7DT is automatically reduced and processed using the \texttt{gpPy-GPU} pipeline \citep{paek2025_gppy_gpu}. 
The \texttt{gpPy-GPU} module accelerates array-based computations via GPU processing, enabling the efficient generation of master bias, dark, and flat frames, as well as their application to the reduction of object images. 
The data reduction procedure includes bias, dark, and flat-field corrections, astrometric calibration using \texttt{SCAMP} \citep{2006ASPC..351..112B} and photometric zeropoint (ZP) calibration using \texttt{SExtractor} \citep{1996A&AS..117..393B}. 
The ZP calibration was derived using isolated, unsaturated sources within the field, matched to synthetic photometry generated from Gaia DR3 XP spectra provided by the Gaia team \citep{2023A&A...674A...1G}. Following the basic instrumental corrections and calibrations, all individual exposures taken in a given filter on the same night were aligned and combined to produce a co-added image. All subsequent source detection and photometric analyses were performed exclusively on these nightly co-added images.

\begin{figure*}
    \includegraphics[width=\linewidth]{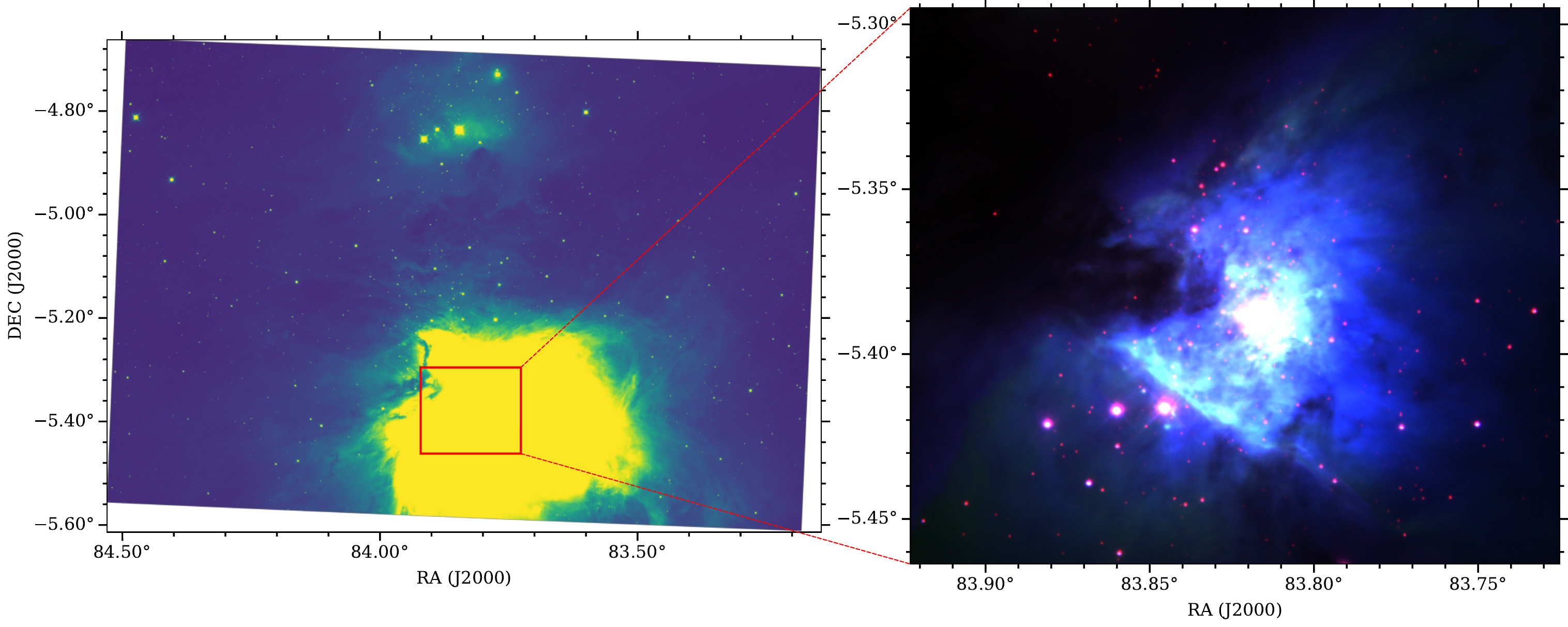}
    \caption{\textbf{Left:} Wide-field image of the Orion A central region obtained with the 7DT telescope using the m650 filter. The image was constructed by combining multiple exposures taken on March 24, 2024. The red square marks the region that is enlarged in the right panel. \textbf{Right:} RGB composite image of the inner region of the Orion Nebula using m850 (red), m650 (green), and m500 (blue) filters. The bright ridge running diagonally from the upper left to the lower right corresponds to the Orion Bar, a photon-dominated interface shaped by the strong ultraviolet (UV) radiation from the nearby Trapezium Cluster. The structure reveals fine filaments, embedded stars, and illuminated gas layers characteristic of this transitional region between the HII region and the molecular cloud.
    \label{fig:rgb}}
\end{figure*}

These observations provided photometric data that contributed to the identification and analysis of variability in YSOs within these regions. The systematic coverage across the optical spectrum with medium-band filters enabled robust characterization of YSO variability mechanisms.
Figure \ref{fig:rgb} presents a full-field image obtained with the m650 filter, alongside an RGB composite of a zoomed-in region encompassing the Orion Bar. These images demonstrate the capability of 7DT to resolve fine-scale structures such as ionization fronts, filaments, and embedded YSOs.

\subsection{Deep Learning-based Automated Detection of Satellite Trails via Structural Similarity}
Satellites in low Earth orbit frequently cross the field of view in wide-field astronomical surveys, leaving linear trails in CCD images. These trails can contaminate photometric measurements and hinder the detection of faint astronomical sources. To mitigate this issue, we implemented an automated satellite trajectory detection pipeline that relies on the Structural Similarity Index Measure (SSIM; \citealt{wang2004image}). By comparing image regions against reference patterns, SSIM effectively captures structural distortions and enables reliable identification of both bright and faint satellite trails, including fragmented ones caused by exposure irregularities.

Building upon this idea, we adopted deep learning to automate the detection of satellite trails. To build the training dataset, we grouped images by observation date and wavelength, in total 29 groups. For our data, each group contains 30 time-consecutive images taken at the same wavelength resulting in 870 images. Each image was manually labeled to indicate the presence or absence of satellite trails, enabling supervised learning. Because the task was framed as a binary classification problem, we balanced the dataset by generating a set of synthetic trails in the images. Specifically, each artificial linear trail was added to a subset of trail-free images so that the ratio of positive (trail-present) and negative (trail-absent) samples was 1:1. With this balanced dataset established, we then developed a method that differs from conventional line detection tasks by leveraging domain-specific knowledge of astronomical imaging.

We leveraged the idea that astronomical observations capture multiple exposures of the same field under nearly identical conditions, such that the background sky remains constant while only stellar variability or satellite trails may differ. Thus, our idea was to capture the pixel-wise difference between each pair of images within the same group, which would effectively highlight the trail-present areas. For this, we employed the SSIM, which measures the similarity between two images based on their luminance, contrast, and structure. Using SSIM, we derived additional feature maps designed to suppress the static astronomical background while enhancing transient features such as satellite trails. These maps effectively isolate transient structures while minimizing contributions from the background.

\begin{figure*}
    \includegraphics[width=\textwidth]{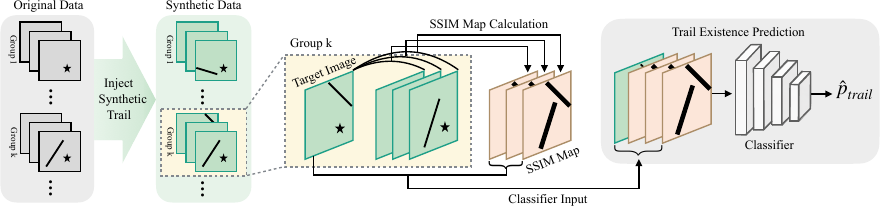}
    \caption{Schematic overview of our SSIM-based deep-learning pipeline for satellite-trail detection. Each group consists of 30 time-consecutive images of the same field at a fixed wavelength (Original Data). For training, synthetic linear trails are injected into trail-free images to construct a balanced positive/negative sample (Synthetic Data). For each target image, we compute SSIM maps between this image and the remaining images in the group, which helps suppress the static astronomical background and enhance transient linear features. The original image and its SSIM maps are then fed into a ResNet-34 classifier, which outputs the probability ($\hat{p}_{trail}$) that a satellite trail is present in the target image. \label{fig:ssim}}
\end{figure*}

Each target image, along with its SSIM maps, was used as the input to our deep learning model, as summarized in Figure \ref{fig:ssim}. The model was based on a ResNet architecture \citep{he2016deep}, consisting of 34 convolutional layers with residual connections, a global average pooling layer, and a fully connected output layer for binary classification. Training was performed using the Adam optimizer with a learning rate of 0.001. A learning rate scheduler (ReduceLROnPlateau) and early stopping were applied, with a maximum of 500 epochs allowed to ensure convergence while avoiding overfitting.

The SSIM-based ResNet classifier achieved an accuracy of 0.97, a recall of 0.92, a precision of 0.94, and an F1-score of 0.93 on an evaluation set of 270 images. We applied the classifier to the full dataset and flagged 241 of 1140 images ($\sim$20\%) as satellite-trail-contaminated. These images were excluded prior to image combination, and the final co-added images used for photometry were constructed only from the remaining trail-free exposures.

\subsection{Photometry}
Multiple exposures per filter were aligned and combined into deep images after removing satellite-trail-contaminated images identified by the SSIM-based ResNet classifier. Weighted averaging with sigma clipping was used to suppress noise. Source detection was performed on each combined image using Source Extractor, with a detection threshold of 3$\sigma$ and a minimum area of 5 connected pixels, i.e., at least five contiguous pixels above the detection threshold. Aperture magnitudes were calibrated with standard stars, and shape parameters such as ellipticity and FWHM were measured for each source.

\begin{figure}
    \includegraphics[width=\columnwidth]{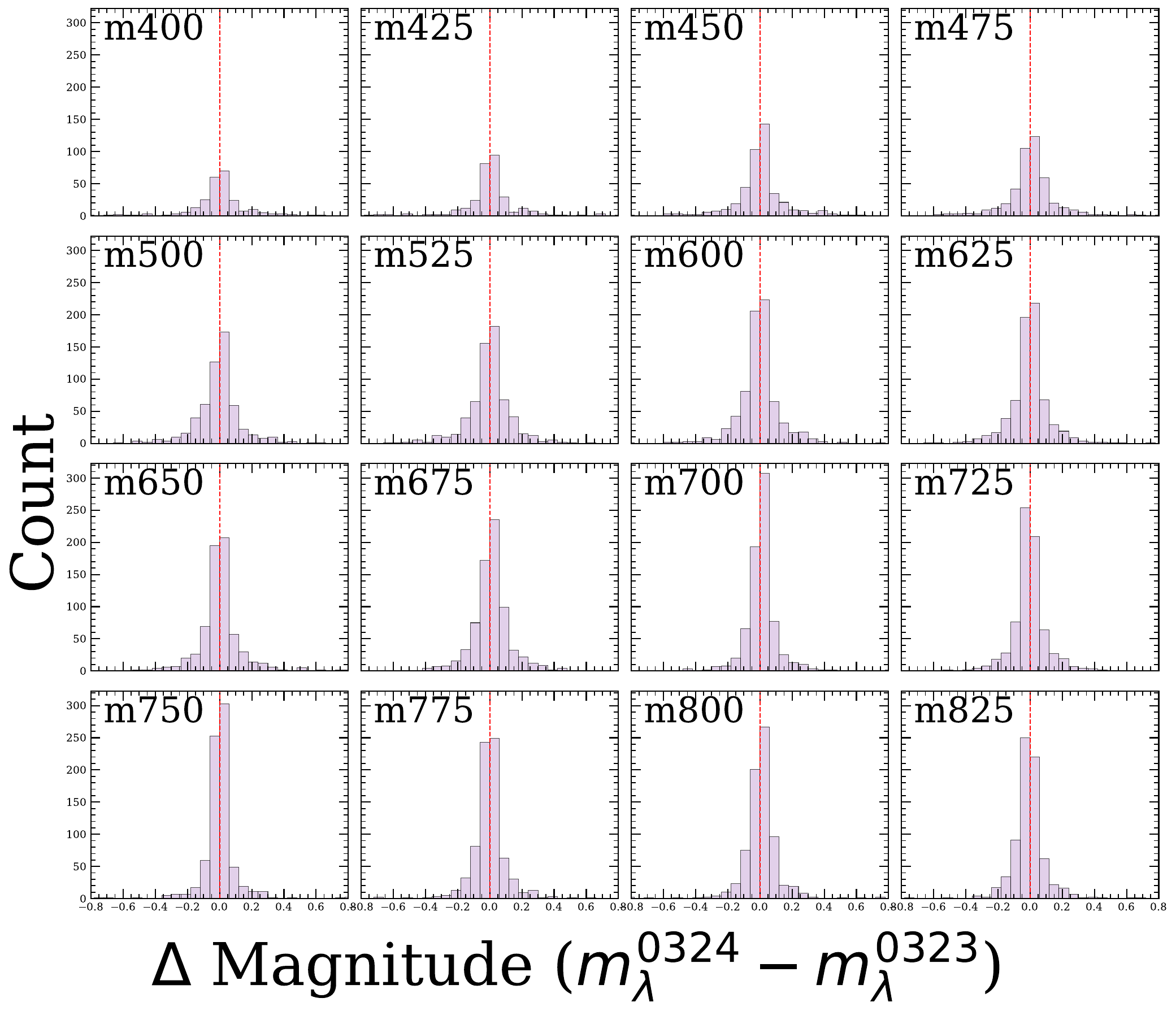}
    \caption{Distributions of two-epoch magnitude differences, $\Delta m_{\lambda}=m_{\lambda}^{0324}-m_{\lambda}^{0323}$, for sources detected on both nights in each of the 16 medium-band filters. The vertical dashed line marks $\Delta m_{\lambda}=0$.}
    \label{fig:all_dmag_hist}
\end{figure}

To refine the photometric zero points for each filter and epoch, we matched bright, unsaturated stars (11--15 mag) in each 7DT image to Gaia counterparts and computed the offset between the observed 7DT magnitudes and Gaia-based synthetic magnitudes \citep{2023gaia}. After rejecting outliers, we adopted the robust (sigma-clipped) offset and applied it as a zero-point correction to all sources. As a sanity check on the inter-night calibration, we examined the distributions of two-epoch magnitude differences $\Delta m_{\lambda}=m_{\lambda}^{0324}-m_{\lambda}^{0323}$ for sources detected on both nights (Figure \ref{fig:all_dmag_hist}). Most filters show distributions that are sharply peaked near $\Delta m_{\lambda}\simeq 0$, which indicates stable relative calibration between the two nights.

To ensure photometric reliability, we selected only sources with extraction flags of zero (i.e., no blending, saturation, or truncation issues) and ellipticity values less than 0.3. We then matched detections across all filters and both nights within a 2.5$''$ radius to construct a unified catalog. For catalog construction, we required detections in at least 10 filters across the two-night dataset (without requiring a detection in the same filter on both nights), yielding 1091 objects. 
To reduce background contamination, we excluded sources with Gaia distances outside 300--600 pc, leaving 892 objects. This broad distance range was adopted to retain sources plausibly associated with the Orion A cloud, whose central region lies at a distance of $\sim$414 pc, while rejecting obvious foreground and background contaminants. We treated this distance-selected sample as our initial YSO-candidate pool and visually inspected their SEDs to remove sources with compromised photometry. A total of 90 objects exhibiting inconsistent or non-physical SEDs were discarded, yielding a final sample of 802 YSO candidates.

Among these, 33 sources were identified as early spectral types (O, B, and A) via SIMBAD. 
Since our analysis focuses on the optical variability of low-mass PMS stars (T Tauri stars), which are typically late-type objects \citep{1989Bertout,2016Hartmann}, we excluded these early-type sources. Intermediate-mass PMS stars, such as Herbig Ae/Be stars, are outside the scope of this study. After this exclusion, the final working sample contains 769 sources.

For reference, we cross-matched the final sample with known YSO catalogs. Of the 769 objects, 412 have counterparts in YSOVAR \citep{2011Morales}, where variability types and periods at 3--4 $\mu$m are reported. In addition, 251 sources are matched to the Spitzer Orion catalog \citep{2012Megeath}, and three have HOPS \citep{2016Furlan} identifications.

Among the 769 objects in our final sample, 726 are cross-matched with known catalogs. The remaining 43 sources, which lack counterparts in HOPS, YSOVAR, or SIMBAD, are retained in this work as YSO candidates. These candidates satisfy our adopted detection, distance, and SED-based selection criteria, and their nature was further assessed through visual inspection of their individual SEDs.
We conservatively treat these 43 objects as candidate YSOs pending further spectroscopic or high-resolution follow-up observations. To assist future identification, Table \ref{tab:list} provides archival cross-identifications, including SIMBAD identifiers when available, and clearly flags the 43 unmatched candidates for subsequent study.
The resulting 7DT photometric catalog of the final working sample of 769 sources is presented in Table \ref{tab:list}. It includes source coordinates, two-night medium-band photometry, catalog cross-identifications, and a variability flag, and the full table is available in machine-readable form.

\begin{deluxetable*}{lcc cccc lcc}
\tabletypesize{\scriptsize}
\tablewidth{0pt}
\tablecaption{7DT Photometric catalog of YSO candidates in Orion A Central Region \label{tab:list}}
\tablehead{
\colhead{ID} & \colhead{RA} & \colhead{DEC} & \multicolumn{4}{c}{Magnitude} & \colhead{Other Name} & \colhead{Var} & \colhead{Stetson $J$}
}
\colnumbers
\startdata
1 & 83.171566 & -5.073494 & m400 & m425 & m450 & m475 & TYC 4774-505-1 & X &  2.412 \\
 & & & 12.69$\pm$0.00 & 12.45$\pm$0.00 & \nodata & \nodata &  & & \\
 & & & 12.69$\pm$0.00 & 12.46$\pm$0.00 & \nodata & \nodata &  & & \\
\cline{4-7}
 & & & m500 & m525 & m600 & m625 &  & & \\
 & & & \nodata & \nodata & 11.83$\pm$0.00 & 11.74$\pm$0.00 &  & & \\
 & & & \nodata & \nodata & 11.85$\pm$0.00 & 11.76$\pm$0.00 &  & & \\
\cline{4-7}
 & & & m650 & m675 & m700 & m725 &  & & \\
 & & & 11.82$\pm$0.00 & 11.74$\pm$0.00 & 11.76$\pm$0.00 & 11.78$\pm$0.00 &  & & \\
 & & & 11.84$\pm$0.00 & 11.78$\pm$0.00 & 11.77$\pm$0.00 & 11.77$\pm$0.00 &  & & \\
\cline{4-7}
 & & & m750 & m775 & m800 & m825 &  & & \\
 & & & 11.73$\pm$0.00 & 11.70$\pm$0.01 & 11.65$\pm$0.01 & 11.65$\pm$0.01 &  & & \\
 & & & 11.74$\pm$0.00 & 11.72$\pm$0.01 & 11.68$\pm$0.01 & 11.69$\pm$0.01 &  & & \\
 2 & 83.177773 & -4.992015 & m400 & m425 & m450 & m475 & V* V926 Ori & X & -0.075 \\
 & & & 16.94$\pm$0.05 & 16.52$\pm$0.04 & \nodata & \nodata &  & & \\
 & & & 16.96$\pm$0.06 & 16.34$\pm$0.04 & \nodata & \nodata &  & & \\
\cline{4-7}
 & & & m500 & m525 & m600 & m625 &  & & \\
 & & & \nodata & \nodata & 14.49$\pm$0.01 & 14.24$\pm$0.01 &  & & \\
 & & & \nodata & \nodata & 14.46$\pm$0.01 & 14.30$\pm$0.01 &  & & \\
\cline{4-7}
 & & & m650 & m675 & m700 & m725 &  & & \\
 & & & 14.18$\pm$0.01 & 14.11$\pm$0.01 & 14.03$\pm$0.01 & 13.93$\pm$0.01 &  & & \\
 & & & 14.18$\pm$0.01 & 14.12$\pm$0.01 & 14.03$\pm$0.01 & 13.90$\pm$0.01 &  & & \\
\cline{4-7}
 & & & m750 & m775 & m800 & m825 &  & & \\
 & & & 13.68$\pm$0.01 & 13.63$\pm$0.01 & 13.53$\pm$0.01 & 13.43$\pm$0.02 &  & & \\
 & & & 13.69$\pm$0.01 & 13.66$\pm$0.01 & 13.51$\pm$0.01 & 13.39$\pm$0.02 &  & & \\
\enddata
\tablecomments{The full Table \ref{tab:list} is published in its entirety in the electronic edition. Columns (2) and (3) list the J2000 coordinates in degrees. Columns (4)--(7) present the 7DT photometric results; for each four-band block, the first data row corresponds to observations on 2024 March 23, and the second data row to observations on 2024 March 24. Column (8) provides cross-identifications from other catalogs, including YSOVAR, HOPS, and 2MASS where available. Column (9) gives the variability flag, and Column (10) lists the calculated Stetson $J$ value.}
\end{deluxetable*}

\section{Analysis and Results}\label{sec:results}
We quantify photometric variability on day timescales by measuring the magnitude difference between the two epochs ($\Delta m_{\lambda} = m_{\lambda}^{0324} - m_{\lambda}^{0323}$) for all sources detected on both nights. A set of day-timescale variable candidates is first established by computing the significance $z_{\lambda}=\Delta m_{\lambda}/\sigma_{\Delta,\lambda}$, where $\sigma_{\Delta,\lambda}=\sqrt{\sigma^{2}_{0324,\lambda}+\sigma^{2}_{0323,\lambda}}$ represents the propagated two-epoch uncertainty. A source is initially flagged as variable if at least 80\% of the available filters satisfy $|z_{\lambda}|>3$, which yields 115 variable candidates among the 769 objects.

As an additional criterion, we computed a two-epoch multi-band Stetson index $J$ \citep{1996Stetson}. The Stetson index is commonly used in multi-epoch light-curve analyses to identify correlated variability from quasi-simultaneous paired observations, such that non-variable sources tend to cluster near $J \sim 0$ while genuinely variable sources tend toward positive values. Here we adapt that framework to our two-night 7DT dataset by replacing the usual time-series pairs with same-night pairs of different optical filters. In this way, the statistic measures whether the two-epoch photometric changes are coherent across multiple filters rather than dominated by random scatter or isolated deviations.

We define a two-epoch multi-band version of the index as
\begin{equation}
J = \frac{1}{N_{\rm pair}} \sum_{k=1}^{N_{\rm pair}} \mathrm{sgn}(P_k)\sqrt{|P_k|},
\end{equation}
where $N_{\rm pair}$ is the total number of valid same-night filter pairs. For a given pair $k$ consisting of two different filters ($i$ and $j$) observed on the same night $t$, the product of the normalized residuals is
\begin{equation}
P_k = \delta_i \delta_j.
\end{equation}
The normalized residuals are computed as
\begin{equation}
\delta_{f,t} = \sqrt{\frac{n_f}{n_f-1}} \frac{m_{f,t}-\bar{m}_f}{\sigma_{f,t}},
\end{equation}
where $n_f$ is the number of valid measurements in that filter, $\bar{m}_f$ is the mean of the available two-epoch measurements in that filter, and $\sigma_{f,t}$ is the photometric uncertainty.

In our two-epoch data, $J$ quantifies the coherence of the observed magnitude changes across different filters. Sources dominated by random scatter are expected to have a $J$ value close to 0, whereas sources with consistent multi-band variability tend to have a positive $J$ value.
We computed $J$ for all 769 sources. Among the 115 selected variable candidates, five have negative $J$ values. Because negative $J$ values indicate that opposite-sign paired residuals dominate under our adopted same-night filter-pairing scheme, these objects do not show coherent variability and were not retained in the final variable sample. We therefore adopt 110 sources as the final day-timescale variable candidates. The corresponding variability flag and Stetson $J$ value are listed in Table \ref{tab:list}.

\begin{figure}
    \includegraphics[width=\columnwidth]{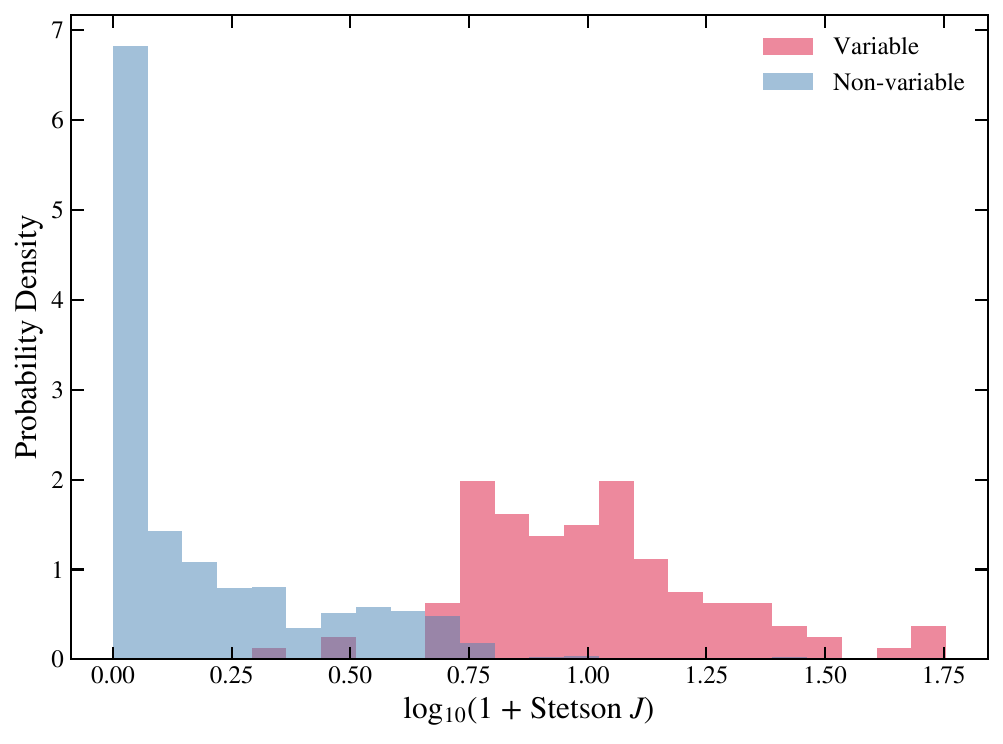}
    \caption{Probability density distribution of the two-epoch multi-band Stetson index $J$ for all 769 sources in our sample, separated into the variable and non-variable groups. The x-axis is shown as $log_{10}(1+J)$ to compress the extended high-$J$ tail and better visualize the separation between the two groups.}
    \label{fig:stetson}
\end{figure}

Figure \ref{fig:stetson} shows the distribution of the resulting $J$ values for 769 sources, separated into the variable and non-variable groups. The variable candidates are systematically shifted toward larger $J$ values, whereas the non-variable sources are concentrated near $J \sim 0$, indicating that our selection preferentially identifies sources with correlated multi-band night-to-night changes rather than isolated deviations in only a few filters. For visualization, the x axis is shown as $log_{10}(1+J)$, which compresses the extended high-$J$ tail and makes the contrast between the sharp low-$J$ peak of the non-variable sources and the broader high-$J$ distribution of the variable sources easier to see. 

\begin{figure}
    \includegraphics[width=\columnwidth]{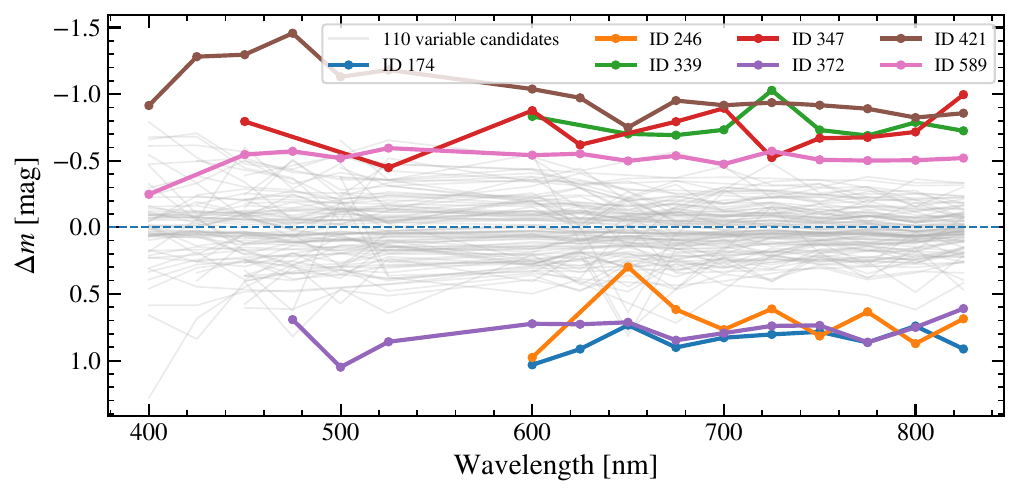}
    \caption{Wavelength dependence of two-epoch magnitude differences for the 110 day-timescale variable candidates. Gray curves show $\Delta m_{\lambda}$ across the 16 medium-band filters. Colored curves highlight the highly variable subset, selected by requiring $|\Delta m_{\lambda}|>0.5$ mag in at least 70\% of the valid filters (with $N_{\rm valid}\ge 9$). The dashed line marks $\Delta m_{\lambda}=0$, where negative values indicate that the source was brighter on March 24, 2024, than on March 23.}
    \label{fig:dm_wave}
\end{figure}

Figure \ref{fig:dm_wave} presents the wavelength dependence of the two-epoch magnitude differences for the 110 variable candidates, summarizing the spectral diversity of day-scale variability. In this plot, gray curves show $\Delta m_{\lambda}$ for all candidates, while colored curves highlight the highly variable subset defined by $|\Delta m_{\lambda}|>0.5$ mag in at least 70\% of the valid filters (with $N_{\rm valid}\ge 9$). These seven highly variable sources also lie in the high-$J$ tail of the Stetson $J$ distribution, with $\log_{10}(1+J)$ values between 1.18 and 1.68, indicating that they are among the most strongly coherent multi-band night-to-night variables in our sample. 
Several objects exhibit pronounced wavelength-dependent trends, where the magnitude change becomes larger toward either the blue or red end of the observed range, whereas others show relatively weak wavelength dependence. This diversity in the $\Delta m(\lambda)$ morphology motivates a uniform, wavelength-dependent characterization of the variability in the following analysis. In addition, different wavelength ranges can be sensitive to different variability mechanisms. For example, hotter accretion-related continuum emission is expected to affect the blue end of the optical range more strongly than the redder bands.

\begin{figure}
    \includegraphics[width=\columnwidth]{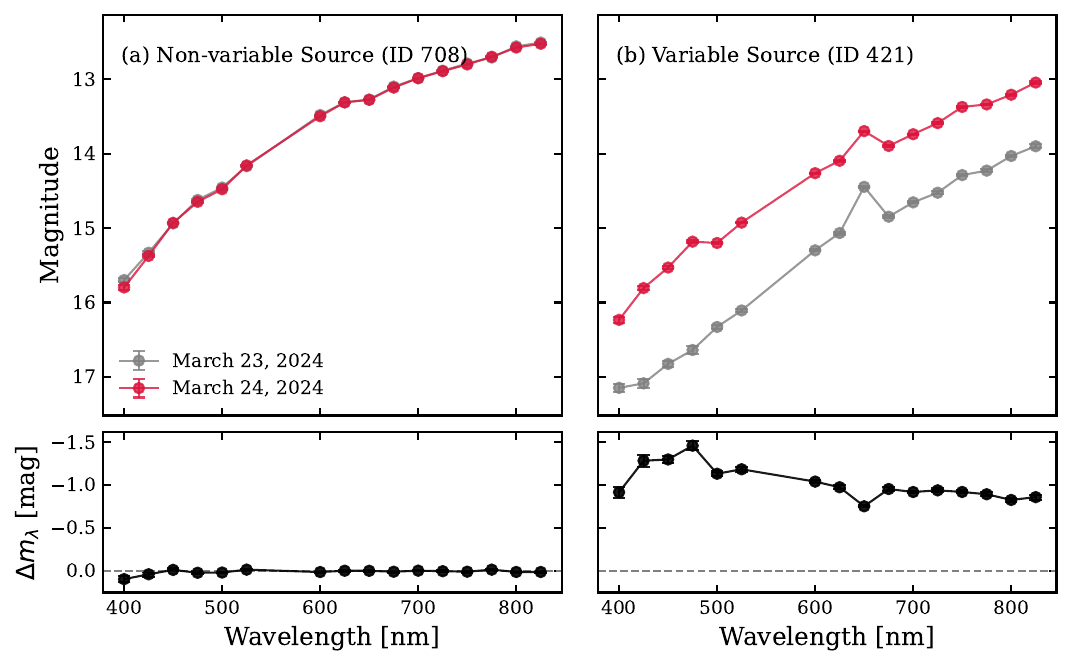}
    \caption{Comparison of a representative source classified as non-variable in our two-night 7DT analysis and a representative variable source. Panel (a) shows ID 708 (2MASS J05365602-0503066), which we classify as non-variable based on the present two-night 7DT data. Panel (b) shows ID 421 (MY Ori), which we classify as variable in our 7DT data and which is also known in the literature as an Orion variable. In both panels, observations from March 23 and March 24, 2024, are denoted by gray and red circles, respectively, and error bars represent 1$\sigma$ photometric uncertainties.}
    \label{fig:sed}
\end{figure}

Figure \ref{fig:sed} compares a non-variable source (ID 708) and a representative variable source (ID 421) using both their two-epoch optical medium-band SEDs and the corresponding band-by-band magnitude differences, $\Delta m_\lambda = m_{0324,\lambda} - m_{0323,\lambda}$. The non-variable source shows $\Delta m_\lambda$ values consistent with zero across the observed bands, whereas the variable object exhibits systematically negative $\Delta m_\lambda$ values that become larger toward shorter wavelengths, indicating stronger brightening on 2024 March 24 at the blue end of the optical range.

Classical T Tauri stars are well known for their irregular variability \citep{1945Joy}, exhibiting changes on timescales ranging from hours to days for routine behavior, weeks to months for bursts, and months to years, in some cases up to decades, for outbursts \citep{2023Fischer}. The corresponding amplitudes range from low-level variations \citep[e.g.,][]{2014Cody,2014Stauffer,2015Venuti}, to several magnitude brightenings in rare eruptive events such as EXor- and FUor-type outbursts \citep[e.g.,][]{2013Lorenzetti,2023Fischer}.

The two-night variability spectra (Figure \ref{fig:dm_wave}) show that many sources exhibit coherent, smoothly varying wavelength dependence across the optical medium bands. To relate these shapes to plausible drivers, we compared five simple templates:
extinction-like changes parameterized with the optical extinction law of Cardelli, Clayton, \& Mathis (\citeyear{1989Cardelli}, hereafter CCM89) using $R_V=3.1$ and $R_V=5.5$, a gray (wavelength-independent) change, and two spot-like toy models (Hot Spot and Cold Spot) represented by a two-temperature surface mixture. 
The goal is a shape-based, empirical classification to identify which simple template best matches the shape of the observed $\Delta m_\lambda$ spectrum, rather than to provide a unique physical inference from two epochs. A nearly gray change, for instance, may reflect weakly wavelength-dependent attenuation, such as partial occultation by optically thick clumps or a warp near the inner disk rim, or extinction dominated by large grains that flattens the optical extinction curve \citep{2015McGinnis,2015Stauffer}. 

For extinction-like changes, we assume 
\begin{equation} 
\Delta m_\lambda = a \, \frac{A(\lambda)}{A_V}, 
\end{equation} 
with $A(\lambda)/A_V$ evaluated at the filter wavelengths for a given $R_V$ (CCM89), and $a$ obtained by a weighted linear fit. 
A positive value ($a>0$) corresponds to increased attenuation (dimming) on March 24, while a negative value ($a<0$) corresponds to decreased attenuation (brightening) on March 24. 
We refer to $a$ as an effective extinction-template scale factor (denoted $\Delta A_{V,\rm eff}$), which would equal a physical $\Delta A_V$ only in the limiting case where the two-epoch change is produced purely by extinction. 
It should be regarded as a template normalization rather than a direct measurement of $\Delta A_V$ derived from a SED/extinction analysis. 
The wavelength-independent (gray) template is simply $\Delta m_\lambda=a_{\rm gray}$.

For spot-like modulation, we adopt a two-component flux template:
\begin{equation}
F_\lambda(f) = (1-f)\,B_\lambda(T_{\rm phot}) + f\,B_\lambda(T_{\rm spot}),
\end{equation}
where $B_\lambda(T)$ is the Planck function at temperature $T$ and $f$ is the spot filling factor on the stellar surface. 
We convert the model flux ratio between the two epochs into a magnitude difference:
\begin{equation}
\Delta m_\lambda = -2.5\log_{10}\left(\frac{F_{\lambda,0324}}{F_{\lambda,0323}}\right).
\end{equation}
We fix the photospheric temperature to $T_{\rm phot}=4000$ K \citep{2013Pecaut} for simplicity and consider two spot cases: (i) Hot Spots with $T_{\rm spot}$ ranging from 4300 to 20000 K ($f \le 0.5$, steps of 0.01), and (ii) Cold Spots with $T_{\rm spot}$ ranging from 2400 to 3800 K ($f \le 0.8$, steps of 0.01).

\begin{figure}
    \centering
    \includegraphics[width=\columnwidth]{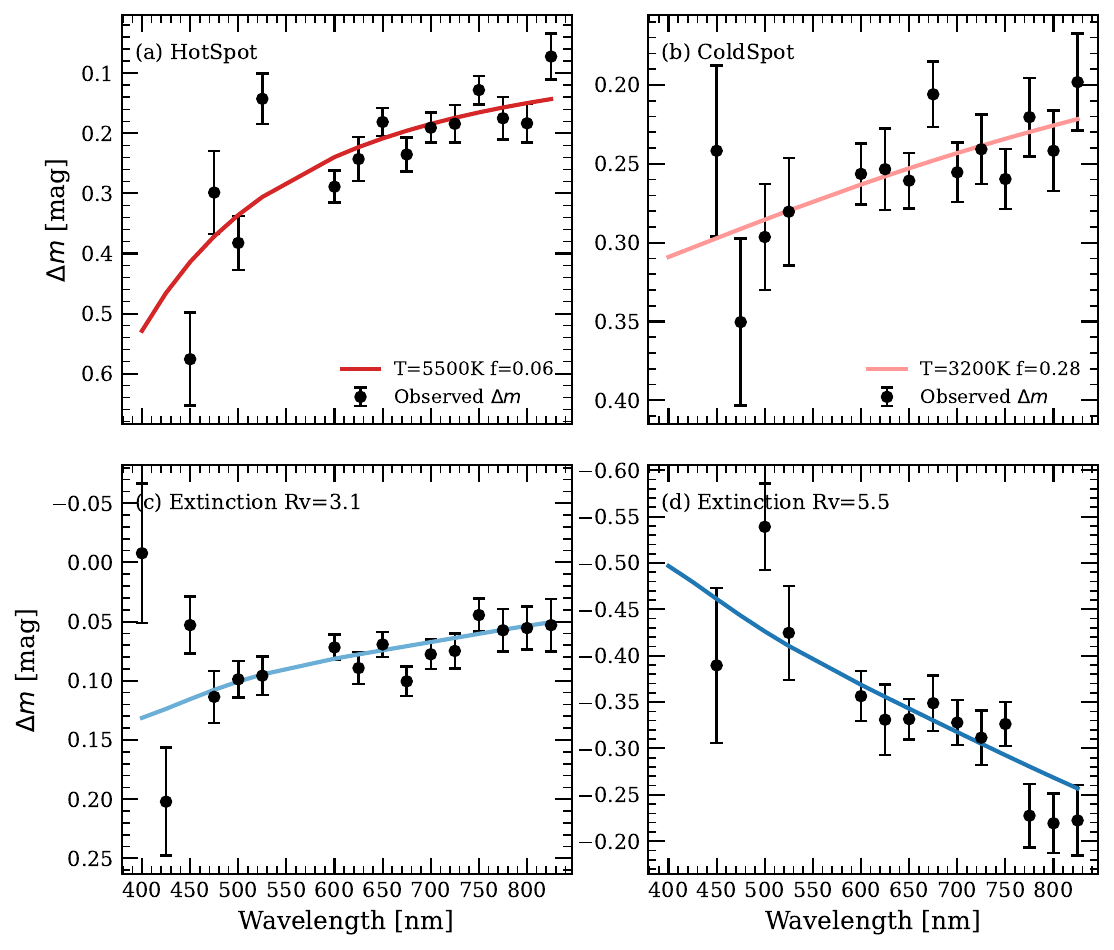}
    \caption{Representative examples of the model fitting in $\Delta m$ space. Each panel shows the observed two-epoch changes with uncertainties and the corresponding best-fit model curve. The four objects are selected to illustrate a Hot Spot, a Cold Spot, and extinction-like solutions for $R_V=3.1$ and $R_V=5.5$.}
    \label{fig:model_examples}
\end{figure}

We assign each variable to the template that minimizes the reduced chi-square, $\chi^2_\nu$, in $\Delta m_\lambda$ space (requiring valid detection in $N_{\rm valid}\ge 8$ bands, with ${\rm dof}=N_{\rm valid}-1$). 
The resulting distribution is dominated by spot-like solutions: Cold Spot is the most frequent template (37/110), followed by Hot Spot (22/110). 
Extinction-like templates account for 37/110 objects in total (split between the two values of $R_V$), while 14/110 are best described by the Gray template. 
Representative examples of the model fitting are shown in Figure \ref{fig:model_examples}.

\section{DISCUSSION} 
Within the classical framework for interpreting this variability \citep[e.g.,][]{1994Herbst}, changes in extinction, magnetospheric accretion, and starspot coverage all contribute and can vary with time. 
Extinction-related variability often manifests as sharp flux drops that can be either short-lived or of longer duration; accretion variability can appear as sharp bursts lasting hours to days, or as smoother changes on timescales of weeks and longer; and spots give rise to flux variations periodically modulated by the stellar rotation \citep{2023Fischer}. 

Recent high-cadence, space-based optical light curve studies show that physical interpretations based on photometry alone are often degenerate. Careful consideration of the light curve morphology, multi-band behavior, and, when available, spectroscopic diagnostics is required to distinguish between variability dominated by accretion, extinction, and spots \citep[e.g.,][]{2014Stauffer,2015Venuti,2017Cody,2023Fischer}.

In this work, we focus on the day-timescale variability of T Tauri stars. On these timescales, the observed changes are likely to reflect a combination of rotational modulation by spots, stochastic fluctuations in the magnetospheric accretion rate, and modest variations in line-of-sight extinction within the inner disk. Our two-epoch 7DT observations, separated by one day, therefore probe the short-timescale regime of the variability spectrum rather than the rare, large-amplitude outbursts that unfold over months to years.

Against this background, we interpret the template-fitting results presented in Section 4 and summarized in Table \ref{tab:modelscan}. Among the template classes, the median variability amplitude, quantified by $\max|\Delta m_\lambda|$ across the 16 medium-band filters, is largest for the Hot Spot and Extinction ($R_V=3.1$) templates and smallest for the Extinction ($R_V=5.5$) template. This is consistent with the expectation that a steeper wavelength dependence can yield larger absolute changes in certain optical bands for a given overall normalization.

The median $\chi^2_\nu$ values indicate that, although these simple templates capture the dominant trends of the two-epoch spectra, they do not provide an adequate description for all sources. 
In particular, the Gray template has a large median $\chi^2_\nu$, and even the spot and extinction templates frequently yield $\chi^2_\nu$ values well above unity. This suggests that a single-component picture is often insufficient on day timescales, plausibly because multiple mechanisms act concurrently (e.g., extinction changes combined with hot components), or because some bands may be influenced by emission-line contributions and modest accretion-related continuum excess (veiling). 
Consequently, the best-fit templates in Table \ref{tab:modelscan} should be interpreted as compact descriptors of the dominant wavelength-dependent trend in $\Delta m_\lambda$, rather than as unique physical explanations for individual objects.

Table \ref{tab:modelscan} also reports the fraction of sources in each template class that satisfy our simple m650 excess criterion (the ``emission-flag fraction''). Specifically, we flag an object when its m650 magnitude is brighter than the local continuum, estimated from neighboring filters such as m625 and m675, by more than a fixed threshold. 
Because the m650 bandpass is designed to cover the H$\alpha$ ($\lambda$6563) emission line, this flag serves as an indirect photometric indicator of excess emission in the H$\alpha$ region. Such an excess may be associated with accretion activity, but it can also be affected by winds and, especially in sources with weak accretion, chromospheric emission \citep{2024Claes,2025Zsidi}. 

For the 110 variables, the emission-flag fraction is 0.455 for the Hot Spot template, 0.432 for Cold Spot, 0.438 for Extinction ($R_V=3.1$), 0.381 for Extinction ($R_V=5.5$), and 0.500 for Gray. Although the Gray category shows the highest m650 excess fraction, these objects also tend to have large reduced $\chi^2_\nu$ values, implying that their two-epoch $\Delta m_\lambda$ spectra are not well captured by a single wavelength-independent component. We therefore interpret the Gray template cautiously, regarding it as a poorly fit subset in which multiple processes or band-dependent contributions may be significant.

When focusing on the non-Gray classes, the Hot Spot template exhibits the highest m650 excess fraction. This is physically plausible within the magnetospheric accretion framework, as an increase in accretion shock luminosity can simultaneously enhance a hot continuum component and strengthen hydrogen line emission, producing both a Hot Spot-like $\Delta m_\lambda$ shape and an m650-band excess. By contrast, the extinction-like and Cold Spot-like templates are expected to be dominated by geometric changes, which do not inherently require a strong m650 excess. We emphasize, however, that our m650 excess criterion is a continuum-referenced photometric proxy. It is sensitive to the adopted local pseudo-continuum estimate and does not uniquely map to an H$\alpha$ equivalent width. 

Complementary diagnostics, such as simultaneous spectroscopy or higher-cadence multi-band monitoring, will be needed to determine whether the m650 excesses in each class primarily trace accretion-related variability or instead reflect more complex, mixed contributions. Although our 7DT wavelength coverage potentially includes additional optical tracers beyond H$\alpha$, the present dataset samples them only through medium-band photometry, and the absence of the m550 and m575 bands in this observation further limits our ability to isolate several line-sensitive regions. We therefore defer a more tracer-specific analysis to future follow-up studies. 

In Figure \ref{fig:dm_wave}, the seven highlighted Orion variable sources demonstrate that the two-night (March 23–24) changes are not merely scatter, but often form coherent, smoothly varying wavelength-dependent patterns across the optical medium bands. 
Most of these objects possess independent variability flags from external time-domain surveys and are classified in YSOVAR as "periodic" or "curved" variables with timescales ranging from several to tens of days (e.g., 8.4, 12, 27, 46, and 88 days; \citealt{2024LeeSE}). 
Additionally, one source (ID 174) is reported as irregular in NEOWISE \citep{2021ParkWS}. 
These external confirmations validate that our two-epoch $\Delta m_\lambda$ spectra provide a meaningful snapshot diagnostic of ongoing variability. 

The diversity of the $\Delta m_\lambda$ shapes is notable: some are nearly gray, while others exhibit gentle slopes with wavelength. 
This diversity is consistent with the well-established picture that multiple drivers, such as weakly wavelength-dependent attenuation from partial occultation, gray extinction, or distinct spot-, accretion-, and extinction-like processes, can operate simultaneously even within the same star-forming region.
Ultimately, the dense optical medium-band sampling provided by 7DT offers a powerful, compact method for characterizing variability shapes. 
When combined with long-baseline survey data from YSOVAR and NEOWISE, this approach can effectively prioritize high-value targets for subsequent spectroscopic follow-up and deeper physical interpretation.

\begin{deluxetable}{l ccccc}
\tabletypesize{\small}
\tablewidth{0pt}
\tablecaption{Summary of the template results for 110 variables \label{tab:modelscan}}
\tablehead{
\colhead{Best template} & \colhead{$N$} & \colhead{median $\chi^2_\nu$} & \colhead{median $\max|\Delta m_\lambda|$} & \colhead{typical parameters} & \colhead{emission-flag fraction}
}
\startdata
Hot Spot & 22 & 7.0 & 0.40 & $T_{\rm spot}\simeq5200$ K, $f \simeq 0.05$ & 0.455 \\
Cold Spot & 37 & 5.6 & 0.27 & $T_{\rm spot}\simeq3200$ K, $f \simeq 0.23$ & 0.432 \\
Extinction ($R_V{=}3.1$) & 16 & 4.5 & 0.37 & $\Delta A_{V,\rm eff}(=a)\simeq-0.04$ mag & 0.438 \\
Extinction ($R_V{=}5.5$) & 21 & 3.7 & 0.20 & $\Delta A_{V,\rm eff}(=a)\simeq-0.03$ mag & 0.381 \\
Gray & 14 & 23.4 & 0.35 & $a_{\rm gray}\simeq+0.13$ mag & 0.500 \\
\enddata
\tablecomments{The variability amplitude is characterized by $\max|\Delta m_\lambda|$ across the 16 medium-band filters (400--825 nm). Parameter values are medians of the best-fit solutions within each template.}
\end{deluxetable}

\section{SUMMARY}
We present a first analysis of day-timescale optical variability in the Orion A central region using two-night 7DT medium-band photometry. 
The 7DT observations provide optical spectral data with 16 medium-band filters spanning 400--825 nm, enabling direct comparisons between the two nights. 

To mitigate contamination from satellite trails, we implemented an automated trail-detection pipeline that combines SSIM-based feature maps with a deep-learning classifier. 
A primary technical contribution of this work is that the resulting ResNet-34 model enables reliable automated rejection of satellite-trail-contaminated exposures. 
On a labeled evaluation set, it achieves an accuracy of 0.97 and an F1-score of 0.93. When applied to the full dataset, it flags 241 of 1140 images($\sim$20\%) as contaminated.
This scalable screening step substantially reduces the need for manual inspection and helps preserve the quality of the final co-added images and photometric measurements.

After removing trail-contaminated exposures identified by this pipeline, we performed photometry and two-epoch variability measurements, selecting a working sample of 769 YSO candidates. 
From this sample, we identified 110 variable objects ($\sim$14\%), including seven highly variable sources with extreme changes of $|\Delta m_\lambda|>0.5$ mag.

To describe the observed $\Delta m_\lambda$ shapes, we compare five simple templates: two extinction-like templates based on the CCM89 law ($R_V=3.1$ and $R_V=5.5$), a gray (wavelength-independent) template, and two spot-like toy models (hot and cold) implemented as two-temperature surface mixtures. 
The best-fit distribution is dominated by spot-like solutions, with Cold Spot (37/110) and Hot Spot (22/110) being most common, while 37/110 sources are best matched by extinction-like templates and 14/110 by the gray template.

We also report, for each template, the fraction of sources that satisfy a simple m650 excess criterion (the emission-flag fraction). 
This flag is based on a photometric excess in m650 relative to a local continuum estimate from neighboring filters, and it should be regarded as an indirect indicator of an m650-band excess rather than a direct equivalent-width measurement. 
Although the m650 excess fraction is highest in the Gray category, we treat this result cautiously because Gray objects tend to show large reduced $\chi^2_\nu$ values, indicating that a wavelength-independent template often provides a poor description. 
Excluding the Gray category, the Hot Spot template shows the highest m650 excess fraction, broadly consistent with the magnetospheric accretion picture in which accretion shocks can contribute both a hot continuum component and enhanced hydrogen line emission.

While the templates describe the dominant trends, many sources show elevated $\chi^2_\nu$ values, particularly for the Gray template, indicating that a single-component model is often insufficient on day timescales. 
This likely reflects a mixture of mechanisms and/or band-dependent contributions that are not explained by continuum-based templates. 

Future monitoring with 7DT spanning not only higher cadence but also a range of cadences and longer time baselines, together with complementary diagnostics, will be important for breaking degeneracies between accretion, extinction, and spot modulation and for clarifying the physical origin of the most complex two-epoch profiles.

\begin{acknowledgments}
This work was supported by the National Research Foundation of Korea (NRF) grant funded by the Korea government (MSIT; grant No. RS-2024-00416859 and RS-2026-25490557). MI acknowledges support from the NRF grant, RS-2026-25490019, of MSIT. G.S.H.P. acknowledges support from the Pan-STARRS project, which is a project of the Institute for Astronomy of the University of Hawai'i, and is supported by the NASA SSO Near Earth Observation Program under grants 80NSSC18K0971, NNX14AM74G, NNX12AR65G, NNX13AQ47G, NNX08AR22G, 80NSSC21K1572, and by the State of Hawai'i. HC acknowledge(s) support from the National Research Foundation of Korea(NRF) grant (RS-2025-00573214) funded by the Korea government(MSIT). SWC acknowledges support from the NRF grants funded by the Ministry of Education (RS-2023-00245013) and the MIST (RS-2026-25489059). DT is supported by the NRF grant, RS-2024-00343729. 
\end{acknowledgments}

\bibliography{references}{}

@ARTICLE{2023gaia,
       author = {{Gaia Collaboration} and {Montegriffo}, P. and {Bellazzini}, M. and {De Angeli}, F. and {Andrae}, R. and {Barstow}, M.~A. and {Bossini}, D. and {Bragaglia}, A. and {Burgess}, P.~W. and {Cacciari}, C. and {Carrasco}, J.~M. and {Chornay}, N. and {Delchambre}, L. and {Evans}, D.~W. and {Fouesneau}, M. and {Fr{\'e}mat}, Y. and {Garabato}, D. and {Jordi}, C. and {Manteiga}, M. and {Massari}, D. and {Palaversa}, L. and {Pancino}, E. and {Riello}, M. and {Ruz Mieres}, D. and {Sanna}, N. and {Santove{\~n}a}, R. and {Sordo}, R. and {Vallenari}, A. and {Walton}, N.~A. and {Brown}, A.~G.~A. and {Prusti}, T. and {de Bruijne}, J.~H.~J. and {Arenou}, F. and {Babusiaux}, C. and {Biermann}, M. and {Creevey}, O.~L. and {Ducourant}, C. and {Eyer}, L. and {Guerra}, R. and {Hutton}, A. and {Klioner}, S.~A. and {Lammers}, U.~L. and {Lindegren}, L. and {Luri}, X. and {Mignard}, F. and {Panem}, C. and {Pourbaix}, D. and {Randich}, S. and {Sartoretti}, P. and {Soubiran}, C. and {Tanga}, P. and {Bailer-Jones}, C.~A.~L. and {Bastian}, U. and {Drimmel}, R. and {Jansen}, F. and {Katz}, D. and {Lattanzi}, M.~G. and {van Leeuwen}, F. and {Bakker}, J. and {Casta{\~n}eda}, J. and {Fabricius}, C. and {Galluccio}, L. and {Guerrier}, A. and {Heiter}, U. and {Masana}, E. and {Messineo}, R. and {Mowlavi}, N. and {Nicolas}, C. and {Nienartowicz}, K. and {Pailler}, F. and {Panuzzo}, P. and {Riclet}, F. and {Roux}, W. and {Seabroke}, G.~M. and {Th{\'e}venin}, F. and {Gracia-Abril}, G. and {Portell}, J. and {Teyssier}, D. and {Altmann}, M. and {Audard}, M. and {Bellas-Velidis}, I. and {Benson}, K. and {Berthier}, J. and {Blomme}, R. and {Busonero}, D. and {Busso}, G. and {C{\'a}novas}, H. and {Carry}, B. and {Cellino}, A. and {Cheek}, N. and {Clementini}, G. and {Damerdji}, Y. and {Davidson}, M. and {de Teodoro}, P. and {Nu{\~n}ez Campos}, M. and {Dell'Oro}, A. and {Esquej}, P. and {Fern{\'a}ndez-Hern{\'a}ndez}, J. and {Fraile}, E. and {Garc{\'\i}a-Lario}, P. and {Gosset}, E. and {Haigron}, R. and {Halbwachs}, J.-L. and {Hambly}, N.~C. and {Harrison}, D.~L. and {Hern{\'a}ndez}, J. and {Hestroffer}, D. and {Hodgkin}, S.~T. and {Holl}, B. and {Jan{\ss}en}, K. and {Jevardat de Fombelle}, G. and {Jordan}, S. and {Krone-Martins}, A. and {Lanzafame}, A.~C. and {L{\"o}ffler}, W. and {Marchal}, O. and {Marrese}, P.~M. and {Moitinho}, A. and {Muinonen}, K. and {Osborne}, P. and {Pauwels}, T. and {Recio-Blanco}, A. and {Reyl{\'e}}, C. and {Rimoldini}, L. and {Roegiers}, T. and {Rybizki}, J. and {Sarro}, L.~M. and {Siopis}, C. and {Smith}, M. and {Sozzetti}, A. and {Utrilla}, E. and {van Leeuwen}, M. and {Abbas}, U. and {{\'A}brah{\'a}m}, P. and {Abreu Aramburu}, A. and {Aerts}, C. and {Aguado}, J.~J. and {Ajaj}, M. and {Aldea-Montero}, F. and {Altavilla}, G. and {{\'A}lvarez}, M.~A. and {Alves}, J. and {Anderson}, R.~I. and {Anglada Varela}, E. and {Antoja}, T. and {Baines}, D. and {Baker}, S.~G. and {Balaguer-N{\'u}{\~n}ez}, L. and {Balbinot}, E. and {Balog}, Z. and {Barache}, C. and {Barbato}, D. and {Barros}, M. and {Bartolom{\'e}}, S. and {Bassilana}, J.-L. and {Bauchet}, N. and {Becciani}, U. and {Berihuete}, A. and {Bernet}, M. and {Bertone}, S. and {Bianchi}, L. and {Binnenfeld}, A. and {Blanco-Cuaresma}, S. and {Boch}, T. and {Bombrun}, A. and {Bouquillon}, S. and {Bramante}, L. and {Breedt}, E. and {Bressan}, A. and {Brouillet}, N. and {Brugaletta}, E. and {Bucciarelli}, B. and {Burlacu}, A. and {Butkevich}, A.~G. and {Buzzi}, R. and {Caffau}, E. and {Cancelliere}, R. and {Cantat-Gaudin}, T. and {Carballo}, R. and {Carlucci}, T. and {Carnerero}, M.~I. and {Casamiquela}, L. and {Castellani}, M. and {Castro-Ginard}, A. and {Chaoul}, L. and {Charlot}, P. and {Chemin}, L. and {Chiaramida}, V. and {Chiavassa}, A. and {Comoretto}, G. and {Contursi}, G. and {Cooper}, W.~J. and {Cornez}, T. and {Cowell}, S. and {Crifo}, F. and {Cropper}, M. and {Crosta}, M. and {Crowley}, C. and {Dafonte}, C. and {Dapergolas}, A.},
        title = "{Gaia Data Release 3. The Galaxy in your preferred colours: Synthetic photometry from Gaia low-resolution spectra}",
      journal = {\aap},
         year = 2023,
        month = jun,
       volume = {674},
          eid = {A33},
        pages = {A33},
          doi = {10.1051/0004-6361/202243709},
archivePrefix = {arXiv},
       eprint = {2206.06215},
 primaryClass = {astro-ph.SR},
       adsurl = {https://ui.adsabs.harvard.edu/abs/2023A&A...674A..33G}
}

@ARTICLE{2006Natta,
       author = {{Natta}, A. and {Testi}, L. and {Randich}, S.},
        title = "{Accretion in the {\ensuremath{\rho}}-Ophiuchi pre-main sequence stars}",
      journal = {\aap},
         year = 2006,
        month = jun,
       volume = {452},
       number = {1},
        pages = {245-252},
          doi = {10.1051/0004-6361:20054706},
archivePrefix = {arXiv},
       eprint = {astro-ph/0602618},
 primaryClass = {astro-ph},
       adsurl = {https://ui.adsabs.harvard.edu/abs/2006A&A...452..245N}
}

@ARTICLE{2024Claes,
       author = {{Claes}, R.~A.~B. and {Campbell-White}, J. and {Manara}, C.~F. and {Frasca}, A. and {Natta}, A. and {Alcal{\'a}}, J.~M. and {Armeni}, A. and {Fang}, M. and {Lovell}, J.~B. and {Stelzer}, B. and {Venuti}, L. and {Wyatt}, M. and {Queitsch}, A.},
        title = "{FitteR for Accretion ProPErties of T Tauri stars (FRAPPE): A new approach to use class III spectra to derive stellar and accretion properties}",
      journal = {\aap},
         year = 2024,
        month = oct,
       volume = {690},
          eid = {A122},
        pages = {A122},
          doi = {10.1051/0004-6361/202450885},
archivePrefix = {arXiv},
       eprint = {2407.11866},
 primaryClass = {astro-ph.SR},
       adsurl = {https://ui.adsabs.harvard.edu/abs/2024A&A...690A.122C}
}

@ARTICLE{2025Zsidi,
       author = {{Zsidi}, Gabriella and {K{\'o}sp{\'a}l}, {\'A}gnes and {{\'A}brah{\'a}m}, P{\'e}ter and {Alecian}, Evelyne and {Alencar}, Silvia H.~P. and {Bouvier}, J{\'e}r{\^o}me and {Hussain}, Gaitee A.~J. and {Manara}, Carlo F. and {Siwak}, Michal and {Szab{\'o}}, R{\'o}bert and {Bora}, Zs{\'o}fia and {Cseh}, Borb{\'a}la and {Kalup}, Csilla and {Kiss}, Csaba and {Kriskovics}, Levente and {Kun}, M{\'a}ria and {P{\'a}l}, Andr{\'a}s and {S{\'o}dor}, {\'A}d{\'a}m and {S{\'a}rneczky}, Kriszti{\'a}n and {Szak{\'a}ts}, R{\'o}bert and {Vida}, Kriszti{\'a}n and {Vink{\'o}}, J{\'o}zsef and {Szab{\'o}}, Z{\'o}fia M.},
        title = "{Short- and long-term variations of the high mass accretion rate classical T Tauri star DR Tau}",
      journal = {\aap},
         year = 2025,
        month = jul,
       volume = {699},
          eid = {A221},
        pages = {A221},
          doi = {10.1051/0004-6361/202449576},
archivePrefix = {arXiv},
       eprint = {2505.07684},
 primaryClass = {astro-ph.SR},
       adsurl = {https://ui.adsabs.harvard.edu/abs/2025A&A...699A.221Z}
}

@ARTICLE{2013Lorenzetti,
       author = {{Lorenzetti}, D. and {Antoniucci}, S. and {Giannini}, T. and {Di Paola}, A. and {Arkharov}, A.~A. and {Larionov}, V.~M.},
        title = "{Interpreting the simultaneous variability of near-IR continuum and line emission in young stellar objects}",
      journal = {\apss},
         year = 2013,
        month = feb,
       volume = {343},
       number = {2},
        pages = {535-539},
          doi = {10.1007/s10509-012-1266-4},
archivePrefix = {arXiv},
       eprint = {1210.0317},
 primaryClass = {astro-ph.SR},
       adsurl = {https://ui.adsabs.harvard.edu/abs/2013Ap&SS.343..535L}
}

@ARTICLE{2023Armeni,
       author = {{Armeni}, A. and {Stelzer}, B. and {Claes}, R.~A.~B. and {Manara}, C.~F. and {Frasca}, A. and {Alcal{\'a}}, J.~M. and {Walter}, F.~M. and {K{\'o}sp{\'a}l}, {\'A}. and {Campbell-White}, J. and {Gangi}, M. and {Mauco}, K. and {Tychoniec}, L.},
        title = "{PENELLOPE. V. The magnetospheric structure and the accretion variability of the classical T Tauri star HM Lup}",
      journal = {\aap},
         year = 2023,
        month = nov,
       volume = {679},
          eid = {A14},
        pages = {A14},
          doi = {10.1051/0004-6361/202347051},
archivePrefix = {arXiv},
       eprint = {2309.10591},
 primaryClass = {astro-ph.SR},
       adsurl = {https://ui.adsabs.harvard.edu/abs/2023A&A...679A..14A}
}

@ARTICLE{2022Claes,
       author = {{Claes}, R.~A.~B. and {Manara}, C.~F. and {Garcia-Lopez}, R. and {Natta}, A. and {Fang}, M. and {Fockter}, Z.~P. and {{\'A}brah{\'a}m}, P. and {Alcal{\'a}}, J.~M. and {Campbell-White}, J. and {Caratti o Garatti}, A. and {Covino}, E. and {Fedele}, D. and {Frasca}, A. and {Gameiro}, J.~F. and {Herczeg}, G.~J. and {K{\'o}sp{\'a}l}, {\'A}. and {Petr-Gotzens}, M.~G. and {Rosotti}, G. and {Venuti}, L. and {Zsidi}, G.},
        title = "{PENELLOPE. III. The peculiar accretion variability of XX Cha and its impact on the observed spread of accretion rates}",
      journal = {\aap},
         year = 2022,
        month = aug,
       volume = {664},
          eid = {L7},
        pages = {L7},
          doi = {10.1051/0004-6361/202244135},
archivePrefix = {arXiv},
       eprint = {2208.01447},
 primaryClass = {astro-ph.SR},
       adsurl = {https://ui.adsabs.harvard.edu/abs/2022A&A...664L...7C}
}

@ARTICLE{2017DaRio,
       author = {{Da Rio}, Nicola and {Tan}, Jonathan C. and {Covey}, Kevin R. and {Cottaar}, Michiel and {Foster}, Jonathan B. and {Cullen}, Nicholas C. and {Tobin}, John and {Kim}, Jinyoung S. and {Meyer}, Michael R. and {Nidever}, David L. and {Stassun}, Keivan G. and {Chojnowski}, S. Drew and {Flaherty}, Kevin M. and {Majewski}, Steven R. and {Skrutskie}, Michael F. and {Zasowski}, Gail and {Pan}, Kaike},
        title = "{IN-SYNC. V. Stellar Kinematics and Dynamics in the Orion A Molecular Cloud}",
      journal = {\apj},
         year = 2017,
        month = aug,
       volume = {845},
       number = {2},
          eid = {105},
        pages = {105},
          doi = {10.3847/1538-4357/aa7a5b},
archivePrefix = {arXiv},
       eprint = {1702.04113},
 primaryClass = {astro-ph.GA},
       adsurl = {https://ui.adsabs.harvard.edu/abs/2017ApJ...845..105D}
}

@ARTICLE{2016DaRio,
       author = {{Da Rio}, Nicola and {Tan}, Jonathan C. and {Covey}, Kevin R. and {Cottaar}, Michiel and {Foster}, Jonathan B. and {Cullen}, Nicholas C. and {Tobin}, John J. and {Kim}, Jinyoung S. and {Meyer}, Michael R. and {Nidever}, David L. and {Stassun}, Keivan G. and {Chojnowski}, S. Drew and {Flaherty}, Kevin M. and {Majewski}, Steve and {Skrutskie}, Michael F. and {Zasowski}, Gail and {Pan}, Kaike},
        title = "{IN-SYNC. IV. The Young Stellar Population in the Orion A Molecular Cloud}",
      journal = {\apj},
         year = 2016,
        month = feb,
       volume = {818},
       number = {1},
          eid = {59},
        pages = {59},
          doi = {10.3847/0004-637X/818/1/59},
archivePrefix = {arXiv},
       eprint = {1511.04147},
 primaryClass = {astro-ph.GA},
       adsurl = {https://ui.adsabs.harvard.edu/abs/2016ApJ...818...59D}
}

@ARTICLE{2021Manara,
       author = {{Manara}, C.~F. and {Frasca}, A. and {Venuti}, L. and {Siwak}, M. and {Herczeg}, G.~J. and {Calvet}, N. and {Hernandez}, J. and {Tychoniec}, {\L}. and {Gangi}, M. and {Alcal{\'a}}, J.~M. and {Boffin}, H.~M.~J. and {Nisini}, B. and {Robberto}, M. and {Briceno}, C. and {Campbell-White}, J. and {Sicilia-Aguilar}, A. and {McGinnis}, P. and {Fedele}, D. and {K{\'o}sp{\'a}l}, {\'A}. and {{\'A}brah{\'a}m}, P. and {Alonso-Santiago}, J. and {Antoniucci}, S. and {Arulanantham}, N. and {Bacciotti}, F. and {Banzatti}, A. and {Beccari}, G. and {Benisty}, M. and {Biazzo}, K. and {Bouvier}, J. and {Cabrit}, S. and {Caratti o Garatti}, A. and {Coffey}, D. and {Covino}, E. and {Dougados}, C. and {Eisl{\"o}ffel}, J. and {Ercolano}, B. and {Espaillat}, C.~C. and {Erkal}, J. and {Facchini}, S. and {Fang}, M. and {Fiorellino}, E. and {Fischer}, W.~J. and {France}, K. and {Gameiro}, J.~F. and {Garcia Lopez}, R. and {Giannini}, T. and {Ginski}, C. and {Grankin}, K. and {G{\"u}nther}, H.~M. and {Hartmann}, L. and {Hillenbrand}, L.~A. and {Hussain}, G.~A.~J. and {James}, M.~M. and {Koutoulaki}, M. and {Lodato}, G. and {Mauc{\'o}}, K. and {Mendigut{\'\i}a}, I. and {Mentel}, R. and {Miotello}, A. and {Oudmaijer}, R.~D. and {Rigliaco}, E. and {Rosotti}, G.~P. and {Sanchis}, E. and {Schneider}, P.~C. and {Spina}, L. and {Stelzer}, B. and {Testi}, L. and {Thanathibodee}, T. and {Vink}, J.~S. and {Walter}, F.~M. and {Williams}, J.~P. and {Zsidi}, G.},
        title = "{PENELLOPE: The ESO data legacy program to complement the Hubble UV Legacy Library of Young Stars (ULLYSES). I. Survey presentation and accretion properties of Orion OB1 and {\ensuremath{\sigma}}-Orionis}",
      journal = {\aap},
         year = 2021,
        month = jun,
       volume = {650},
          eid = {A196},
        pages = {A196},
          doi = {10.1051/0004-6361/202140639},
archivePrefix = {arXiv},
       eprint = {2103.12446},
 primaryClass = {astro-ph.SR},
       adsurl = {https://ui.adsabs.harvard.edu/abs/2021A&A...650A.196M}
}

@ARTICLE{2025Tofflemire,
       author = {{Tofflemire}, Benjamin M. and {Manara}, Carlo F. and {Banzatti}, Andrea and {Pontoppidan}, Klaus M. and {Najita}, Joan and {Nisini}, Brunella and {Whelan}, Emma T. and {Campbell-White}, Justyn and {Alqubelat}, Hala and {Kraus}, Adam L. and {Rab}, Christian and {Houge}, Adrien and {Krijt}, Sebastiaan and {Muzerolle}, James and {Fiorellino}, Eleonora and {Benisty}, Myriam and {Tychoniec}, Lukasz and {Salyk}, Colette and {Bourdarot}, Guillaume and {Hyden}, Jacob},
        title = "{Coordinated Space- and Ground-based Monitoring of Accretion Bursts in a Protoplanetary Disk: Establishing Mid-infrared Hydrogen Lines as Accretion Diagnostics for JWST/MIRI}",
      journal = {\apj},
         year = 2025,
        month = jun,
       volume = {985},
       number = {2},
          eid = {224},
        pages = {224},
          doi = {10.3847/1538-4357/adcc23},
archivePrefix = {arXiv},
       eprint = {2504.08029},
 primaryClass = {astro-ph.SR},
       adsurl = {https://ui.adsabs.harvard.edu/abs/2025ApJ...985..224T}
}

@ARTICLE{2024Rogers,
       author = {{Rogers}, Ciar{\'a}n and {de Marchi}, Guido and {Brandl}, Bernhard},
        title = "{Determining stellar accretion rates from Pa$_{{\ensuremath{\alpha}}}$ and Br$_{{\ensuremath{\beta}}}$ emission lines with JWST NIRSpec. Accretion of pre-main-sequence stars in NGC 3603}",
      journal = {\aap},
         year = 2024,
        month = apr,
       volume = {684},
          eid = {L8},
        pages = {L8},
          doi = {10.1051/0004-6361/202449282},
archivePrefix = {arXiv},
       eprint = {2403.09568},
 primaryClass = {astro-ph.SR},
       adsurl = {https://ui.adsabs.harvard.edu/abs/2024A&A...684L...8R}
}

@ARTICLE{2025Fiorellino,
       author = {{Fiorellino}, E. and {Alcal{\'a}}, J.~M. and {Manara}, C.~F. and {Pittman}, C.~V. and {{\'A}brah{\'a}m}, P. and {Venuti}, L. and {Cabrit}, S. and {Claes}, R. and {Fang}, M. and {K{\'o}sp{\'a}l}, {\'A}. and {Lodato}, G. and {Mauco}, K. and {Tychoniec}, {\L}.},
        title = "{PENELLOPE: VII. Revisiting empirical relations to measure accretion luminosity}",
      journal = {\aap},
         year = 2025,
        month = dec,
       volume = {704},
          eid = {A42},
        pages = {A42},
          doi = {10.1051/0004-6361/202556603},
archivePrefix = {arXiv},
       eprint = {2509.21078},
 primaryClass = {astro-ph.SR},
       adsurl = {https://ui.adsabs.harvard.edu/abs/2025A&A...704A..42F}
}

@ARTICLE{2020Komarova,
       author = {{Komarova}, Olena and {Fischer}, William J.},
        title = "{Calibration of Brackett Alpha as an Accretion Indicator in T Tauri Stars}",
      journal = {Research Notes of the American Astronomical Society},
         year = 2020,
        month = jan,
       volume = {4},
       number = {1},
          eid = {6},
        pages = {6},
          doi = {10.3847/2515-5172/ab67bb},
       adsurl = {https://ui.adsabs.harvard.edu/abs/2020RNAAS...4....6K}
}

@ARTICLE{2013Salyk,
       author = {{Salyk}, Colette and {Herczeg}, Gregory J. and {Brown}, Joanna M. and {Blake}, Geoffrey A. and {Pontoppidan}, Klaus M. and {van Dishoeck}, Ewine F.},
        title = "{Measuring Protoplanetary Disk Accretion with H I Pfund {\ensuremath{\beta}}}",
      journal = {\apj},
         year = 2013,
        month = may,
       volume = {769},
       number = {1},
          eid = {21},
        pages = {21},
          doi = {10.1088/0004-637X/769/1/21},
archivePrefix = {arXiv},
       eprint = {1303.4804},
 primaryClass = {astro-ph.SR},
       adsurl = {https://ui.adsabs.harvard.edu/abs/2013ApJ...769...21S}
}

@ARTICLE{2015Rigliaco,
       author = {{Rigliaco}, Elisabetta and {Pascucci}, I. and {Duchene}, G. and {Edwards}, S. and {Ardila}, D.~R. and {Grady}, C. and {Mendigut{\'\i}a}, I. and {Montesinos}, B. and {Mulders}, G.~D. and {Najita}, J.~R. and {Carpenter}, J. and {Furlan}, E. and {Gorti}, U. and {Meijerink}, R. and {Meyer}, M.~R.},
        title = "{Probing Stellar Accretion with Mid-infrared Hydrogen Lines}",
      journal = {\apj},
         year = 2015,
        month = mar,
       volume = {801},
       number = {1},
          eid = {31},
        pages = {31},
          doi = {10.1088/0004-637X/801/1/31},
archivePrefix = {arXiv},
       eprint = {1501.06210},
 primaryClass = {astro-ph.SR},
       adsurl = {https://ui.adsabs.harvard.edu/abs/2015ApJ...801...31R}
}

@ARTICLE{2017Alcala,
       author = {{Alcal{\'a}}, J.~M. and {Manara}, C.~F. and {Natta}, A. and {Frasca}, A. and {Testi}, L. and {Nisini}, B. and {Stelzer}, B. and {Williams}, J.~P. and {Antoniucci}, S. and {Biazzo}, K. and {Covino}, E. and {Esposito}, M. and {Getman}, F. and {Rigliaco}, E.},
        title = "{X-shooter spectroscopy of young stellar objects in Lupus. Accretion properties of class II and transitional objects}",
      journal = {\aap},
         year = 2017,
        month = apr,
       volume = {600},
          eid = {A20},
        pages = {A20},
          doi = {10.1051/0004-6361/201629929},
archivePrefix = {arXiv},
       eprint = {1612.07054},
 primaryClass = {astro-ph.SR},
       adsurl = {https://ui.adsabs.harvard.edu/abs/2017A&A...600A..20A}
}

@ARTICLE{2014Alcala,
       author = {{Alcal{\'a}}, J.~M. and {Natta}, A. and {Manara}, C.~F. and {Spezzi}, L. and {Stelzer}, B. and {Frasca}, A. and {Biazzo}, K. and {Covino}, E. and {Randich}, S. and {Rigliaco}, E. and {Testi}, L. and {Comer{\'o}n}, F. and {Cupani}, G. and {D'Elia}, V.},
        title = "{X-shooter spectroscopy of young stellar objects. IV. Accretion in low-mass stars and substellar objects in Lupus}",
      journal = {\aap},
         year = 2014,
        month = jan,
       volume = {561},
          eid = {A2},
        pages = {A2},
          doi = {10.1051/0004-6361/201322254},
archivePrefix = {arXiv},
       eprint = {1310.2069},
 primaryClass = {astro-ph.SR},
       adsurl = {https://ui.adsabs.harvard.edu/abs/2014A&A...561A...2A}
}

@ARTICLE{2024Boyden,
       author = {{Boyden}, Ryan D. and {Eisner}, Josh A.},
        title = "{Constraining Free{\textendash}Free Emission and Photoevaporative Mass-loss Rates for Known Proplyds and New VLA{\textendash}identified Candidate Proplyds in NGC 1977}",
      journal = {\apj},
         year = 2024,
        month = jun,
       volume = {967},
       number = {2},
          eid = {103},
        pages = {103},
          doi = {10.3847/1538-4357/ad3cd5},
archivePrefix = {arXiv},
       eprint = {2404.04437},
 primaryClass = {astro-ph.EP},
       adsurl = {https://ui.adsabs.harvard.edu/abs/2024ApJ...967..103B}
}

@ARTICLE{1987Bally,
       author = {{Bally}, John and {Langer}, William D. and {Stark}, Antony A. and {Wilson}, Robert W.},
        title = "{Filamentary Structure in the Orion Molecular Cloud}",
      journal = {\apjl},
         year = 1987,
        month = jan,
       volume = {312},
        pages = {L45},
          doi = {10.1086/184817},
       adsurl = {https://ui.adsabs.harvard.edu/abs/1987ApJ...312L..45B}
}

@ARTICLE{1996Stetson,
       author = {{Stetson}, Peter B.},
        title = "{On the Automatic Determination of Light-Curve Parameters for Cepheid Variables}",
      journal = {\pasp},
         year = 1996,
        month = oct,
       volume = {108},
        pages = {851},
          doi = {10.1086/133808},
       adsurl = {https://ui.adsabs.harvard.edu/abs/1996PASP..108..851S}
}

@ARTICLE{2013Pecaut,
       author = {{Pecaut}, Mark J. and {Mamajek}, Eric E.},
        title = "{Intrinsic Colors, Temperatures, and Bolometric Corrections of Pre-main-sequence Stars}",
      journal = {\apjs},
         year = 2013,
        month = sep,
       volume = {208},
       number = {1},
          eid = {9},
        pages = {9},
          doi = {10.1088/0067-0049/208/1/9},
archivePrefix = {arXiv},
       eprint = {1307.2657},
 primaryClass = {astro-ph.SR},
       adsurl = {https://ui.adsabs.harvard.edu/abs/2013ApJS..208....9P}
}

@ARTICLE{2016Furlan,
       author = {{Furlan}, E. and {Fischer}, W.~J. and {Ali}, B. and {Stutz}, A.~M. and {Stanke}, T. and {Tobin}, J.~J. and {Megeath}, S.~T. and {Osorio}, M. and {Hartmann}, L. and {Calvet}, N. and {Poteet}, C.~A. and {Booker}, J. and {Manoj}, P. and {Watson}, D.~M. and {Allen}, L.},
        title = "{The Herschel Orion Protostar Survey: Spectral Energy Distributions and Fits Using a Grid of Protostellar Models}",
      journal = {\apjs},
         year = 2016,
        month = may,
       volume = {224},
       number = {1},
          eid = {5},
        pages = {5},
          doi = {10.3847/0067-0049/224/1/5},
archivePrefix = {arXiv},
       eprint = {1602.07314},
 primaryClass = {astro-ph.SR},
       adsurl = {https://ui.adsabs.harvard.edu/abs/2016ApJS..224....5F}
}

@ARTICLE{2015Stauffer,
       author = {{Stauffer}, John and {Cody}, Ann Marie and {McGinnis}, Pauline and {Rebull}, Luisa and {Hillenbrand}, Lynne A. and {Turner}, Neal J. and {Carpenter}, John and {Plavchan}, Peter and {Carey}, Sean and {Terebey}, Susan and {Morales-Calder{\'o}n}, Mar{\'\i}a and {Alencar}, Silvia H.~P. and {Bouvier}, Jerome and {Venuti}, Laura and {Hartmann}, Lee and {Calvet}, Nuria and {Micela}, Giusi and {Flaccomio}, Ettore and {Song}, Inseok and {Gutermuth}, Rob and {Barrado}, David and {Vrba}, Frederick J. and {Covey}, Kevin and {Padgett}, Debbie and {Herbst}, William and {Gillen}, Edward and {Lyra}, Wladimir and {Medeiros Guimaraes}, Marcelo and {Bouy}, Herve and {Favata}, Fabio},
        title = "{CSI 2264: Characterizing Young Stars in NGC 2264 With Short-Duration Periodic Flux Dips in Their Light Curves}",
      journal = {\aj},
         year = 2015,
        month = apr,
       volume = {149},
       number = {4},
          eid = {130},
        pages = {130},
          doi = {10.1088/0004-6256/149/4/130},
archivePrefix = {arXiv},
       eprint = {1501.06609},
 primaryClass = {astro-ph.SR},
       adsurl = {https://ui.adsabs.harvard.edu/abs/2015AJ....149..130S}
}

@ARTICLE{2015McGinnis,
       author = {{McGinnis}, P.~T. and {Alencar}, S.~H.~P. and {Guimar{\~a}es}, M.~M. and {Sousa}, A.~P. and {Stauffer}, J. and {Bouvier}, J. and {Rebull}, L. and {Fonseca}, N.~N.~J. and {Venuti}, L. and {Hillenbrand}, L. and {Cody}, A.~M. and {Teixeira}, P.~S. and {Aigrain}, S. and {Favata}, F. and {F{\H{u}}r{\'e}sz}, G. and {Vrba}, F.~J. and {Flaccomio}, E. and {Turner}, N.~J. and {Gameiro}, J.~F. and {Dougados}, C. and {Herbst}, W. and {Morales-Calder{\'o}n}, M. and {Micela}, G.},
        title = "{CSI 2264: Probing the inner disks of AA Tauri-like systems in NGC 2264}",
      journal = {\aap},
         year = 2015,
        month = may,
       volume = {577},
          eid = {A11},
        pages = {A11},
          doi = {10.1051/0004-6361/201425475},
archivePrefix = {arXiv},
       eprint = {1502.07692},
 primaryClass = {astro-ph.SR},
       adsurl = {https://ui.adsabs.harvard.edu/abs/2015A&A...577A..11M}
}

@inproceedings{2024ChoiHH,
author = {Hyeonho Choi and Myungshin Im and Ji Hoon Kim},
title = {{TCSpy: Multitelescope array control software for 7-Dimensional Telescope (7DT)}},
volume = {13101},
booktitle = {Software and Cyberinfrastructure for Astronomy VIII},
editor = {Jorge Ibsen and Gianluca Chiozzi},
organization = {International Society for Optics and Photonics},
publisher = {SPIE},
pages = {131012V},
year = {2024},
doi = {10.1117/12.3018636},
URL = {https://doi.org/10.1117/12.3018636}
}

@INPROCEEDINGS{2024Im,
       author = {{Im}, Myungshin},
        title = "{7-Dimensional Telescope for Multi-Messenger Astronomy}",
    booktitle = {45th COSPAR Scientific Assembly},
         year = 2024,
       volume = {45},
        month = jul,
        pages = {1744},
       adsurl = {https://ui.adsabs.harvard.edu/abs/2024cosp...45.1744I}
}

@ARTICLE{2018Schneider,
       author = {{Schneider}, P.~C. and {Manara}, C.~F. and {Facchini}, S. and {G{\"u}nther}, H.~M. and {Herczeg}, G.~J. and {Fedele}, D. and {Teixeira}, P.~S.},
        title = "{Multi-epoch monitoring of the AA Tauri-like star V 354 Mon. Indications for a low gas-to-dust ratio in the inner disk warp}",
      journal = {\aap},
         year = 2018,
        month = jun,
       volume = {614},
          eid = {A108},
        pages = {A108},
          doi = {10.1051/0004-6361/201731959},
archivePrefix = {arXiv},
       eprint = {1803.00614},
 primaryClass = {astro-ph.SR},
       adsurl = {https://ui.adsabs.harvard.edu/abs/2018A&A...614A.108S}
}

@ARTICLE{2024Wendeborn,
       author = {{Wendeborn}, John and {Espaillat}, Catherine C. and {Thanathibodee}, Thanawuth and {Robinson}, Connor E. and {Pittman}, Caeley V. and {Calvet}, Nuria and {K{\'o}sp{\'a}l}, {\'A}gnes and {Grankin}, Konstantin N. and {Walter}, Fredrick M. and {Guo}, Zhen and {Eisl{\"o}ffel}, Jochen},
        title = "{A Multiwavelength, Multiepoch Monitoring Campaign of Accretion Variability in T Tauri Stars from the ODYSSEUS Survey. II. Photometric Light Curves}",
      journal = {\apj},
         year = 2024,
        month = aug,
       volume = {971},
       number = {1},
          eid = {96},
        pages = {96},
          doi = {10.3847/1538-4357/ad543d},
archivePrefix = {arXiv},
       eprint = {2405.21071},
 primaryClass = {astro-ph.SR},
       adsurl = {https://ui.adsabs.harvard.edu/abs/2024ApJ...971...96W}
}

@ARTICLE{1998Calvet,
       author = {{Calvet}, Nuria and {Gullbring}, Erik},
        title = "{The Structure and Emission of the Accretion Shock in T Tauri Stars}",
      journal = {\apj},
         year = 1998,
        month = dec,
       volume = {509},
       number = {2},
        pages = {802-818},
          doi = {10.1086/306527},
       adsurl = {https://ui.adsabs.harvard.edu/abs/1998ApJ...509..802C}
}

@misc{paek2025_gppy_gpu,
  author       = {Paek, Gregory S. H.},
  title        = {gppy-gpu v1.0.0: A GPU-Accelerated Imaging Pipeline for the 7DT/7DS},
  year         = {2025},
  publisher    = {Zenodo},
  version      = {v1.0.0},
  doi          = {10.5281/zenodo. 17065902},
  url          = {https://doi.org/10.5281/zenodo. 17065902},
  note         = {Computer software}
}

@ARTICLE{2023A&A...674A...1G,
       author = {{Gaia Collaboration} and {Vallenari}, A. and {Brown}, A.~G.~A. and {Prusti}, T. and {de Bruijne}, J.~H.~J. and {Arenou}, F. and {Babusiaux}, C. and {Biermann}, M. and {Creevey}, O.~L. and {Ducourant}, C. and {Evans}, D.~W. and {Eyer}, L. and {Guerra}, R. and {Hutton}, A. and {Jordi}, C. and {Klioner}, S.~A. and {Lammers}, U.~L. and {Lindegren}, L. and {Luri}, X. and {Mignard}, F. and {Panem}, C. and {Pourbaix}, D. and {Randich}, S. and {Sartoretti}, P. and {Soubiran}, C. and {Tanga}, P. and {Walton}, N.~A. and {Bailer-Jones}, C.~A.~L. and {Bastian}, U. and {Drimmel}, R. and {Jansen}, F. and {Katz}, D. and {Lattanzi}, M.~G. and {van Leeuwen}, F. and {Bakker}, J. and {Cacciari}, C. and {Casta{\~n}eda}, J. and {De Angeli}, F. and {Fabricius}, C. and {Fouesneau}, M. and {Fr{\'e}mat}, Y. and {Galluccio}, L. and {Guerrier}, A. and {Heiter}, U. and {Masana}, E. and {Messineo}, R. and {Mowlavi}, N. and {Nicolas}, C. and {Nienartowicz}, K. and {Pailler}, F. and {Panuzzo}, P. and {Riclet}, F. and {Roux}, W. and {Seabroke}, G.~M. and {Sordo}, R. and {Th{\'e}venin}, F. and {Gracia-Abril}, G. and {Portell}, J. and {Teyssier}, D. and {Altmann}, M. and {Andrae}, R. and {Audard}, M. and {Bellas-Velidis}, I. and {Benson}, K. and {Berthier}, J. and {Blomme}, R. and {Burgess}, P.~W. and {Busonero}, D. and {Busso}, G. and {C{\'a}novas}, H. and {Carry}, B. and {Cellino}, A. and {Cheek}, N. and {Clementini}, G. and {Damerdji}, Y. and {Davidson}, M. and {de Teodoro}, P. and {Nu{\~n}ez Campos}, M. and {Delchambre}, L. and {Dell'Oro}, A. and {Esquej}, P. and {Fern{\'a}ndez-Hern{\'a}ndez}, J. and {Fraile}, E. and {Garabato}, D. and {Garc{\'\i}a-Lario}, P. and {Gosset}, E. and {Haigron}, R. and {Halbwachs}, J.-L. and {Hambly}, N.~C. and {Harrison}, D.~L. and {Hern{\'a}ndez}, J. and {Hestroffer}, D. and {Hodgkin}, S.~T. and {Holl}, B. and {Jan{\ss}en}, K. and {Jevardat de Fombelle}, G. and {Jordan}, S. and {Krone-Martins}, A. and {Lanzafame}, A.~C. and {L{\"o}ffler}, W. and {Marchal}, O. and {Marrese}, P.~M. and {Moitinho}, A. and {Muinonen}, K. and {Osborne}, P. and {Pancino}, E. and {Pauwels}, T. and {Recio-Blanco}, A. and {Reyl{\'e}}, C. and {Riello}, M. and {Rimoldini}, L. and {Roegiers}, T. and {Rybizki}, J. and {Sarro}, L.~M. and {Siopis}, C. and {Smith}, M. and {Sozzetti}, A. and {Utrilla}, E. and {van Leeuwen}, M. and {Abbas}, U. and {{\'A}brah{\'a}m}, P. and {Abreu Aramburu}, A. and {Aerts}, C. and {Aguado}, J.~J. and {Ajaj}, M. and {Aldea-Montero}, F. and {Altavilla}, G. and {{\'A}lvarez}, M.~A. and {Alves}, J. and {Anders}, F. and {Anderson}, R.~I. and {Anglada Varela}, E. and {Antoja}, T. and {Baines}, D. and {Baker}, S.~G. and {Balaguer-N{\'u}{\~n}ez}, L. and {Balbinot}, E. and {Balog}, Z. and {Barache}, C. and {Barbato}, D. and {Barros}, M. and {Barstow}, M.~A. and {Bartolom{\'e}}, S. and {Bassilana}, J.-L. and {Bauchet}, N. and {Becciani}, U. and {Bellazzini}, M. and {Berihuete}, A. and {Bernet}, M. and {Bertone}, S. and {Bianchi}, L. and {Binnenfeld}, A. and {Blanco-Cuaresma}, S. and {Blazere}, A. and {Boch}, T. and {Bombrun}, A. and {Bossini}, D. and {Bouquillon}, S. and {Bragaglia}, A. and {Bramante}, L. and {Breedt}, E. and {Bressan}, A. and {Brouillet}, N. and {Brugaletta}, E. and {Bucciarelli}, B. and {Burlacu}, A. and {Butkevich}, A.~G. and {Buzzi}, R. and {Caffau}, E. and {Cancelliere}, R. and {Cantat-Gaudin}, T. and {Carballo}, R. and {Carlucci}, T. and {Carnerero}, M.~I. and {Carrasco}, J.~M. and {Casamiquela}, L. and {Castellani}, M. and {Castro-Ginard}, A. and {Chaoul}, L. and {Charlot}, P. and {Chemin}, L. and {Chiaramida}, V. and {Chiavassa}, A. and {Chornay}, N. and {Comoretto}, G. and {Contursi}, G. and {Cooper}, W.~J. and {Cornez}, T. and {Cowell}, S. and {Crifo}, F. and {Cropper}, M. and {Crosta}, M. and {Crowley}, C. and {Dafonte}, C. and {Dapergolas}, A. and {David}, M. and {David}, P. and {de Laverny}, P. and {De Luise}, F. and {De March}, R.},
        title = "{Gaia Data Release 3. Summary of the content and survey properties}",
      journal = {\aap},
         year = 2023,
        month = jun,
       volume = {674},
          eid = {A1},
        pages = {A1},
          doi = {10.1051/0004-6361/202243940},
archivePrefix = {arXiv},
       eprint = {2208.00211},
 primaryClass = {astro-ph.GA},
       adsurl = {https://ui.adsabs.harvard.edu/abs/2023A&A...674A...1G}
}

@INPROCEEDINGS{2006ASPC..351..112B,
       author = {{Bertin}, E.},
        title = "{Automatic Astrometric and Photometric Calibration with SCAMP}",
    booktitle = {Astronomical Data Analysis Software and Systems XV},
         year = 2006,
       editor = {{Gabriel}, C. and {Arviset}, C. and {Ponz}, D. and {Enrique}, S.},
       series = {Astronomical Society of the Pacific Conference Series},
       volume = {351},
        month = jul,
        pages = {112},
       adsurl = {https://ui.adsabs.harvard.edu/abs/2006ASPC..351..112B}
}

@ARTICLE{1996A&AS..117..393B,
       author = {{Bertin}, E. and {Arnouts}, S.},
        title = "{SExtractor: Software for source extraction.}",
      journal = {\aaps},
         year = 1996,
        month = jun,
       volume = {117},
        pages = {393-404},
          doi = {10.1051/aas:1996164},
       adsurl = {https://ui.adsabs.harvard.edu/abs/1996A&AS..117..393B}
}

@INCOLLECTION{2008Peterson,
       author = {{Peterson}, D.~E. and {Megeath}, S.~T.},
        title = "{The Orion Molecular Cloud 2/3 and NGC 1977 Regions}",
    booktitle = {Handbook of Star Forming Regions, Volume I},
         year = 2008,
       editor = {{Reipurth}, B.},
       volume = {4},
        pages = {590},
          doi = {10.48550/arXiv.0809.4006},
       adsurl = {https://ui.adsabs.harvard.edu/abs/2008hsf1.book..590P}
}

@ARTICLE{1997Chini,
       author = {{Chini}, R. and {Reipurth}, Bo and {Ward-Thompson}, D. and {Bally}, J. and {Nyman}, L.-{\r{A}}. and {Sievers}, A. and {Billawala}, Y.},
        title = "{Dust Filaments and Star Formation in OMC-2 and OMC-3}",
      journal = {\apjl},
         year = 1997,
        month = jan,
       volume = {474},
       number = {2},
        pages = {L135-L138},
          doi = {10.1086/310436},
       adsurl = {https://ui.adsabs.harvard.edu/abs/1997ApJ...474L.135C}
}

@ARTICLE{2014Venuti,
       author = {{Venuti}, L. and {Bouvier}, J. and {Flaccomio}, E. and {Alencar}, S.~H.~P. and {Irwin}, J. and {Stauffer}, J.~R. and {Cody}, A.~M. and {Teixeira}, P.~S. and {Sousa}, A.~P. and {Micela}, G. and {Cuillandre}, J.-C. and {Peres}, G.},
        title = "{Mapping accretion and its variability in the young open cluster NGC 2264: a study based on u-band photometry}",
      journal = {\aap},
         year = 2014,
        month = oct,
       volume = {570},
          eid = {A82},
        pages = {A82},
          doi = {10.1051/0004-6361/201423776},
archivePrefix = {arXiv},
       eprint = {1408.0432},
 primaryClass = {astro-ph.SR},
       adsurl = {https://ui.adsabs.harvard.edu/abs/2014A&A...570A..82V}
}

@ARTICLE{2015Venuti,
       author = {{Venuti}, L. and {Bouvier}, J. and {Irwin}, J. and {Stauffer}, J.~R. and {Hillenbrand}, L.~A. and {Rebull}, L.~M. and {Cody}, A.~M. and {Alencar}, S.~H.~P. and {Micela}, G. and {Flaccomio}, E. and {Peres}, G.},
        title = "{UV variability and accretion dynamics in the young open cluster NGC 2264}",
      journal = {\aap},
         year = 2015,
        month = sep,
       volume = {581},
          eid = {A66},
        pages = {A66},
          doi = {10.1051/0004-6361/201526164},
archivePrefix = {arXiv},
       eprint = {1506.06858},
 primaryClass = {astro-ph.SR},
       adsurl = {https://ui.adsabs.harvard.edu/abs/2015A&A...581A..66V}
}

@inproceedings{he2016deep,
 title = {Deep Residual Learning for Image Recognition},
 author = {He, Kaiming and Zhang, Xiangyu and Ren, Shaoqing and Sun, Jian},
 booktitle = {Proceedings of the IEEE Conference on Computer Vision and Pattern Recognition (CVPR)},
 pages = {770--778},
 year = {2016},
 doi = {10.1109/CVPR.2016.90}
}

@article{wang2004image,
 title = {Image quality assessment: From error visibility to structural similarity},
 author = {Wang, Zhou and Bovik, Alan C. and Sheikh, Hamid R. and Simoncelli, Eero P.},
 journal = {IEEE Transactions on Image Processing},
 volume = {13},
 number = {4},
 pages = {600--612},
 year = {2004},
 publisher = {IEEE},
 doi = {10.1109/TIP.2003.819861}
}

@ARTICLE{2017Kounkel,
       author = {{Kounkel}, Marina and {Hartmann}, Lee and {Loinard}, Laurent and {Ortiz-Le{\'o}n}, Gisela N. and {Mioduszewski}, Amy J. and {Rodr{\'\i}guez}, Luis F. and {Dzib}, Sergio A. and {Torres}, Rosa M. and {Pech}, Gerardo and {Galli}, Phillip A.~B. and {Rivera}, Juana L. and {Boden}, Andrew F. and {Evans}, II, Neal J. and {Brice{\~n}o}, Cesar and {Tobin}, John J.},
        title = "{The Gould{\textquoteright}s Belt Distances Survey (GOBELINS) II. Distances and Structure toward the Orion Molecular Clouds}",
      journal = {\apj},
         year = 2017,
        month = jan,
       volume = {834},
       number = {2},
          eid = {142},
        pages = {142},
          doi = {10.3847/1538-4357/834/2/142},
archivePrefix = {arXiv},
       eprint = {1609.04041},
 primaryClass = {astro-ph.SR},
       adsurl = {https://ui.adsabs.harvard.edu/abs/2017ApJ...834..142K}
}

@ARTICLE{1945Joy,
       author = {{Joy}, Alfred H.},
        title = "{T Tauri Variable Stars.}",
      journal = {\apj},
         year = 1945,
        month = sep,
       volume = {102},
        pages = {168},
          doi = {10.1086/144749},
       adsurl = {https://ui.adsabs.harvard.edu/abs/1945ApJ...102..168J}
}

@ARTICLE{1989Cardelli,
       author = {{Cardelli}, Jason A. and {Clayton}, Geoffrey C. and {Mathis}, John S.},
        title = "{The Relationship between Infrared, Optical, and Ultraviolet Extinction}",
      journal = {\apj},
         year = 1989,
        month = oct,
       volume = {345},
        pages = {245},
          doi = {10.1086/167900},
       adsurl = {https://ui.adsabs.harvard.edu/abs/1989ApJ...345..245C}
}

@ARTICLE{2000Aso,
       author = {{Aso}, Yoshiyuki and {Tatematsu}, Ken'ichi and {Sekimoto}, Yutaro and {Nakano}, Takenori and {Umemoto}, Tomofumi and {Koyama}, Katsuji and {Yamamoto}, Satoshi},
        title = "{Dense Cores and Molecular Outflows in the OMC-2/3 Region}",
      journal = {\apjs},
         year = 2000,
        month = dec,
       volume = {131},
       number = {2},
        pages = {465-482},
          doi = {10.1086/317378},
       adsurl = {https://ui.adsabs.harvard.edu/abs/2000ApJS..131..465A}
}

@ARTICLE{2021Espaillat,
       author = {{Espaillat}, C.~C. and {Robinson}, C.~E. and {Romanova}, M.~M. and {Thanathibodee}, T. and {Wendeborn}, J. and {Calvet}, N. and {Reynolds}, M. and {Muzerolle}, J.},
        title = "{Measuring the density structure of an accretion hot spot}",
      journal = {\nat},
         year = 2021,
        month = sep,
       volume = {597},
       number = {7874},
        pages = {41-44},
          doi = {10.1038/s41586-021-03751-5},
archivePrefix = {arXiv},
       eprint = {2109.00510},
 primaryClass = {astro-ph.SR},
       adsurl = {https://ui.adsabs.harvard.edu/abs/2021Natur.597...41E}
}

@ARTICLE{2012Alencar,
       author = {{Alencar}, S.~H.~P. and {Bouvier}, J. and {Walter}, F.~M. and {Dougados}, C. and {Donati}, J. -F. and {Kurosawa}, R. and {Romanova}, M. and {Bonfils}, X. and {Lima}, G.~H.~R.~A. and {Massaro}, S. and {Ibrahimov}, M. and {Poretti}, E.},
        title = "{Accretion dynamics in the classical T Tauri star V2129 Ophiuchi}",
      journal = {\aap},
         year = 2012,
        month = may,
       volume = {541},
          eid = {A116},
        pages = {A116},
          doi = {10.1051/0004-6361/201118395},
archivePrefix = {arXiv},
       eprint = {1203.6331},
 primaryClass = {astro-ph.SR},
       adsurl = {https://ui.adsabs.harvard.edu/abs/2012A&A...541A.116A}
}

@ARTICLE{1998Muzerolle,
       author = {{Muzerolle}, James and {Calvet}, Nuria and {Hartmann}, Lee},
        title = "{Magnetospheric Accretion Models for the Hydrogen Emission Lines of T Tauri Stars}",
      journal = {\apj},
         year = 1998,
        month = jan,
       volume = {492},
       number = {2},
        pages = {743-753},
          doi = {10.1086/305069},
       adsurl = {https://ui.adsabs.harvard.edu/abs/1998ApJ...492..743M}
}

@ARTICLE{2001Muzerolle,
       author = {{Muzerolle}, James and {Calvet}, Nuria and {Hartmann}, Lee},
        title = "{Emission-Line Diagnostics of T Tauri Magnetospheric Accretion. II. Improved Model Tests and Insights into Accretion Physics}",
      journal = {\apj},
         year = 2001,
        month = apr,
       volume = {550},
       number = {2},
        pages = {944-961},
          doi = {10.1086/319779},
       adsurl = {https://ui.adsabs.harvard.edu/abs/2001ApJ...550..944M}
}

@ARTICLE{1994Hartmann,
       author = {{Hartmann}, Lee and {Hewett}, Robert and {Calvet}, Nuria},
        title = "{Magnetospheric Accretion Models for T Tauri Stars. I. Balmer Line Profiles without Rotation}",
      journal = {\apj},
         year = 1994,
        month = may,
       volume = {426},
        pages = {669},
          doi = {10.1086/174104},
       adsurl = {https://ui.adsabs.harvard.edu/abs/1994ApJ...426..669H}
}

@ARTICLE{1991Koenigl,
       author = {{Koenigl}, Arieh},
        title = "{Disk Accretion onto Magnetic T Tauri Stars}",
      journal = {\apjl},
         year = 1991,
        month = mar,
       volume = {370},
        pages = {L39},
          doi = {10.1086/185972},
       adsurl = {https://ui.adsabs.harvard.edu/abs/1991ApJ...370L..39K}
}

@ARTICLE{1988Walter,
       author = {{Walter}, Frederick M. and {Brown}, A. and {Mathieu}, R.~D. and {Myers}, P.~C. and {Vrba}, F.~J.},
        title = "{X-Ray Sources in Regions of Star Formation. III. Naked T Tauri Stars Associated with the Taurus-Auriga Complex}",
      journal = {\aj},
         year = 1988,
        month = jul,
       volume = {96},
        pages = {297},
          doi = {10.1086/114809},
       adsurl = {https://ui.adsabs.harvard.edu/abs/1988AJ.....96..297W}
}

@ARTICLE{1989Bertout,
       author = {{Bertout}, Claude},
        title = "{T Tauri stars: wild as dust.}",
      journal = {\araa},
         year = 1989,
        month = jan,
       volume = {27},
        pages = {351-395},
          doi = {10.1146/annurev.aa.27.090189.002031},
       adsurl = {https://ui.adsabs.harvard.edu/abs/1989ARA&A..27..351B}
}

@INPROCEEDINGS{2024SPIE13094E..0XK,
       author = {{Kim}, Ji Hoon and {Im}, Myungshin and {Lee}, Hyungmok and {Chang}, Seo-Won and {Choi}, Hyeonho and {Paek}, Gregory S.~H.},
        title = "{Introduction to the 7-Dimensional Telescope: commissioning procedures and data characteristics}",
    booktitle = {Ground-based and Airborne Telescopes X},
         year = 2024,
       editor = {{Marshall}, Heather K. and {Spyromilio}, Jason and {Usuda}, Tomonori},
       series = {Society of Photo-Optical Instrumentation Engineers (SPIE) Conference Series},
       volume = {13094},
        month = aug,
          eid = {130940X},
        pages = {130940X},
          doi = {10.1117/12.3019546},
       adsurl = {https://ui.adsabs.harvard.edu/abs/2024SPIE13094E..0XK}
}

@INPROCEEDINGS{2024SPIE13099E..1ZK,
       author = {{Kim}, Ji Hoon and {Im}, Myungshin and {Lee}, Hyungmok and {Chang}, Seo-Won},
        title = "{Project management for ground-based telescope array development}",
    booktitle = {Modeling, Systems Engineering, and Project Management for Astronomy XI},
         year = 2024,
       editor = {{Egner}, S{\'e}bastien E. and {Roberts}, Scott},
       series = {Society of Photo-Optical Instrumentation Engineers (SPIE) Conference Series},
       volume = {13099},
        month = aug,
          eid = {130991Z},
        pages = {130991Z},
          doi = {10.1117/12.3019770},
       adsurl = {https://ui.adsabs.harvard.edu/abs/2024SPIE13099E..1ZK}
}

@ARTICLE{2024LeeSE,
       author = {{Lee}, Sieun and {Lee}, Jeong-Eun and {Contreras Pe{\~n}a}, Carlos and {Johnstone}, Doug and {Herczeg}, Gregory and {Lee}, Seonjae},
        title = "{Mid-infrared Variability of Young Stellar Objects on Timescales of Days to Years}",
      journal = {\apj},
         year = 2024,
        month = feb,
       volume = {962},
       number = {1},
          eid = {38},
        pages = {38},
          doi = {10.3847/1538-4357/ad14f8},
archivePrefix = {arXiv},
       eprint = {2312.05753},
 primaryClass = {astro-ph.SR},
       adsurl = {https://ui.adsabs.harvard.edu/abs/2024ApJ...962...38L}
}

@ARTICLE{2021ParkWS,
       author = {{Park}, Wooseok and {Lee}, Jeong-Eun and {Contreras Pe{\~n}a}, Carlos and {Johnstone}, Doug and {Herczeg}, Gregory and {Lee}, Sieun and {Lee}, Seonjae and {Bhardwaj}, Anupam and {Moriarty-Schieven}, Gerald H.},
        title = "{Quantifying Variability of Young Stellar Objects in the Mid-infrared Over 6 Years with the Near-Earth Object Wide-field Infrared Survey Explorer}",
      journal = {\apj},
         year = 2021,
        month = oct,
       volume = {920},
       number = {2},
          eid = {132},
        pages = {132},
          doi = {10.3847/1538-4357/ac1745},
archivePrefix = {arXiv},
       eprint = {2107.10751},
 primaryClass = {astro-ph.SR},
       adsurl = {https://ui.adsabs.harvard.edu/abs/2021ApJ...920..132P}
}

@INPROCEEDINGS{2023Fischer,
       author = {{Fischer}, W.~J. and {Hillenbrand}, L.~A. and {Herczeg}, G.~J. and {Johnstone}, D. and {Kospal}, A. and {Dunham}, M.~M.},
        title = "{Accretion Variability as a Guide to Stellar Mass Assembly}",
    booktitle = {Protostars and Planets VII},
         year = 2023,
       editor = {{Inutsuka}, S. and {Aikawa}, Y. and {Muto}, T. and {Tomida}, K. and {Tamura}, M.},
       series = {Astronomical Society of the Pacific Conference Series},
       volume = {534},
        month = jul,
        pages = {355},
          doi = {10.48550/arXiv.2203.11257},
archivePrefix = {arXiv},
       eprint = {2203.11257},
 primaryClass = {astro-ph.SR},
       adsurl = {https://ui.adsabs.harvard.edu/abs/2023ASPC..534..355F}
}

@ARTICLE{2011Morales,
       author = {{Morales-Calder{\'o}n}, M. and {Stauffer}, J.~R. and {Hillenbrand}, L.~A. and {Gutermuth}, R. and {Song}, I. and {Rebull}, L.~M. and {Plavchan}, P. and {Carpenter}, J.~M. and {Whitney}, B.~A. and {Covey}, K. and {Alves de Oliveira}, C. and {Winston}, E. and {McCaughrean}, M.~J. and {Bouvier}, J. and {Guieu}, S. and {Vrba}, F.~J. and {Holtzman}, J. and {Marchis}, F. and {Hora}, J.~L. and {Wasserman}, L.~H. and {Terebey}, S. and {Megeath}, T. and {Guinan}, E. and {Forbrich}, J. and {Hu{\'e}lamo}, N. and {Riviere-Marichalar}, P. and {Barrado}, D. and {Stapelfeldt}, K. and {Hern{\'a}ndez}, J. and {Allen}, L.~E. and {Ardila}, D.~R. and {Bayo}, A. and {Favata}, F. and {James}, D. and {Werner}, M. and {Wood}, K.},
        title = "{Ysovar: The First Sensitive, Wide-area, Mid-infrared Photometric Monitoring of the Orion Nebula Cluster}",
      journal = {\apj},
         year = 2011,
        month = may,
       volume = {733},
       number = {1},
          eid = {50},
        pages = {50},
          doi = {10.1088/0004-637X/733/1/50},
archivePrefix = {arXiv},
       eprint = {1103.5238},
 primaryClass = {astro-ph.SR},
       adsurl = {https://ui.adsabs.harvard.edu/abs/2011ApJ...733...50M}
}

@ARTICLE{2019Akimoto,
       author = {{Akimoto}, Hinako and {Itoh}, Yoichi},
        title = "{Optical Spectroscopic Monitoring Observations of a T Tauri Star V409 Tau}",
      journal = {International Journal of Astronomy and Astrophysics},
         year = 2019,
        month = jan,
       volume = {9},
       number = {3},
        pages = {321-334},
          doi = {10.4236/ijaa.2019.93023},
archivePrefix = {arXiv},
       eprint = {1909.09416},
 primaryClass = {astro-ph.SR},
       adsurl = {https://ui.adsabs.harvard.edu/abs/2019IJAA....9..321A}
}

@ARTICLE{2017Rigon,
       author = {{Rigon}, Laura and {Scholz}, Alexander and {Anderson}, David and {West}, Richard},
        title = "{Long-term variability of T Tauri stars using WASP}",
      journal = {\mnras},
         year = 2017,
        month = mar,
       volume = {465},
       number = {4},
        pages = {3889-3901},
          doi = {10.1093/mnras/stw2977},
archivePrefix = {arXiv},
       eprint = {1611.03013},
 primaryClass = {astro-ph.SR},
       adsurl = {https://ui.adsabs.harvard.edu/abs/2017MNRAS.465.3889R}
}

@ARTICLE{2016Hartmann,
       author = {{Hartmann}, Lee and {Herczeg}, Gregory and {Calvet}, Nuria},
        title = "{Accretion onto Pre-Main-Sequence Stars}",
      journal = {\araa},
         year = 2016,
        month = sep,
       volume = {54},
        pages = {135-180},
          doi = {10.1146/annurev-astro-081915-023347},
       adsurl = {https://ui.adsabs.harvard.edu/abs/2016ARA&A..54..135H}
}

@ARTICLE{2012Megeath,
       author = {{Megeath}, S.~T. and {Gutermuth}, R. and {Muzerolle}, J. and {Kryukova}, E. and {Flaherty}, K. and {Hora}, J.~L. and {Allen}, L.~E. and {Hartmann}, L. and {Myers}, P.~C. and {Pipher}, J.~L. and {Stauffer}, J. and {Young}, E.~T. and {Fazio}, G.~G.},
        title = "{The Spitzer Space Telescope Survey of the Orion A and B Molecular Clouds. I. A Census of Dusty Young Stellar Objects and a Study of Their Mid-infrared Variability}",
      journal = {\aj},
         year = 2012,
        month = dec,
       volume = {144},
       number = {6},
          eid = {192},
        pages = {192},
          doi = {10.1088/0004-6256/144/6/192},
archivePrefix = {arXiv},
       eprint = {1209.3826},
 primaryClass = {astro-ph.GA},
       adsurl = {https://ui.adsabs.harvard.edu/abs/2012AJ....144..192M}
}

@ARTICLE{2001Carpenter,
       author = {{Carpenter}, John M. and {Hillenbrand}, Lynne A. and {Skrutskie}, M.~F.},
        title = "{Near-Infrared Photometric Variability of Stars toward the Orion A Molecular Cloud}",
      journal = {\aj},
         year = 2001,
        month = jun,
       volume = {121},
       number = {6},
        pages = {3160-3190},
          doi = {10.1086/321086},
archivePrefix = {arXiv},
       eprint = {astro-ph/0102446},
 primaryClass = {astro-ph},
       adsurl = {https://ui.adsabs.harvard.edu/abs/2001AJ....121.3160C}
}

@ARTICLE{1994Herbst,
       author = {{Herbst}, W. and {Herbst}, D.~K. and {Grossman}, E.~J. and {Weinstein}, D.},
        title = "{Catalogue of UBVRI Photometry of T Tauri Stars and Analysis of the Causes of Their Variability}",
      journal = {\aj},
         year = 1994,
        month = nov,
       volume = {108},
        pages = {1906},
          doi = {10.1086/117204},
       adsurl = {https://ui.adsabs.harvard.edu/abs/1994AJ....108.1906H}
}

@ARTICLE{2017Cody,
       author = {{Cody}, Ann Marie and {Hillenbrand}, Lynne A. and {David}, Trevor J. and {Carpenter}, John M. and {Everett}, Mark E. and {Howell}, Steve B.},
        title = "{A Continuum of Accretion Burst Behavior in Young Stars Observed by K2}",
      journal = {\apj},
         year = 2017,
        month = feb,
       volume = {836},
       number = {1},
          eid = {41},
        pages = {41},
          doi = {10.3847/1538-4357/836/1/41},
archivePrefix = {arXiv},
       eprint = {1612.05599},
 primaryClass = {astro-ph.SR},
       adsurl = {https://ui.adsabs.harvard.edu/abs/2017ApJ...836...41C}
}

@ARTICLE{2021Froebrich,
       author = {{Froebrich}, Dirk and {Derezea}, Efthymia and {Scholz}, Aleks and {Eisl{\"o}ffel}, Jochen and {Vanaverbeke}, Siegfried and {Kume}, Alfred and {Herbert}, Carys and {Campbell-White}, Justyn and {Miller}, Niall and {Stecklum}, Bringfried and {Makin}, Sally V. and {Urtly}, Thomas and {Sold{\'a}n Alfaro}, Francisco C. and {Schwendeman}, Erik and {Stone}, Geoffrey and {Phillips}, Mark and {Fleming}, George and {Gonzalez Farf{\'a}n}, Rafael and {Vanmunster}, Tonny and {Heald}, Michael A. and {Fern{\'a}ndez Ma{\~n}anes}, Esteban and {Nelson}, Tim and {Eggenstein}, Heinz-Bernd and {Dubois}, Franky and {Logie}, Ludwig and {Rau}, Steve and {Wiersema}, Klaas and {Quinn}, Nick and {Rodriguez}, Diego and {Castillo Garc{\'\i}a}, Rafael and {Killestein}, Thomas and {Vale}, Tony and {Licchelli}, Domenico and {Deldem}, Marc and {Piehler}, Georg and {Mo{\'z}dzierski}, Dawid and {Kotysz}, Krzysztof and {Kowalska}, Katarzyna and {Miko{\l}ajczyk}, Przemys{\l}aw and {Futcher}, Stephen R.~L. and {Long}, Timothy P. and {Morales Aimar}, Mario and {Merrikin}, Barry and {Johnstone}, Stephen and {Dubovsk{\'y}}, Pavol A. and {Kudzej}, Igor and {Pickard}, Roger and {Billington}, Samuel J. and {Dover}, Lord and {Zegmott}, Tarik and {Evitts}, Jack J. and {Traspas Munia}, Alejandra and {Price}, Mark C.},
        title = "{A survey for variable young stars with small telescopes - IV. Rotation periods of YSOs in IC 5070}",
      journal = {\mnras},
         year = 2021,
        month = oct,
       volume = {506},
       number = {4},
        pages = {5989-6000},
          doi = {10.1093/mnras/stab2082},
archivePrefix = {arXiv},
       eprint = {2107.08524},
 primaryClass = {astro-ph.SR},
       adsurl = {https://ui.adsabs.harvard.edu/abs/2021MNRAS.506.5989F}
}

@ARTICLE{2014Stauffer,
       author = {{Stauffer}, John and {Cody}, Ann Marie and {Baglin}, Annie and {Alencar}, Silvia and {Rebull}, Luisa and {Hillenbrand}, Lynne A. and {Venuti}, Laura and {Turner}, Neal J. and {Carpenter}, John and {Plavchan}, Peter and {Findeisen}, Krzysztof and {Carey}, Sean and {Terebey}, Susan and {Morales-Calder{\'o}n}, Mar{\'\i}a and {Bouvier}, Jerome and {Micela}, Giusi and {Flaccomio}, Ettore and {Song}, Inseok and {Gutermuth}, Rob and {Hartmann}, Lee and {Calvet}, Nuria and {Whitney}, Barbara and {Barrado}, David and {Vrba}, Frederick J. and {Covey}, Kevin and {Herbst}, William and {Furesz}, Gabor and {Aigrain}, Suzanne and {Favata}, Fabio},
        title = "{CSI 2264: Characterizing Accretion-burst Dominated Light Curves for Young Stars in NGC 2264}",
      journal = {\aj},
         year = 2014,
        month = apr,
       volume = {147},
       number = {4},
          eid = {83},
        pages = {83},
          doi = {10.1088/0004-6256/147/4/83},
archivePrefix = {arXiv},
       eprint = {1401.6600},
 primaryClass = {astro-ph.SR},
       adsurl = {https://ui.adsabs.harvard.edu/abs/2014AJ....147...83S}
}

@ARTICLE{2016Bozhinova,
       author = {{Bozhinova}, I. and {Scholz}, A. and {Eisl{\"o}ffel}, J.},
        title = "{Variability in young very low mass stars: two surprises from spectrophotometric monitoring}",
      journal = {\mnras},
         year = 2016,
        month = may,
       volume = {458},
       number = {3},
        pages = {3118-3133},
          doi = {10.1093/mnras/stw455},
archivePrefix = {arXiv},
       eprint = {1602.07925},
 primaryClass = {astro-ph.SR},
       adsurl = {https://ui.adsabs.harvard.edu/abs/2016MNRAS.458.3118B}
}

@INPROCEEDINGS{2007Bouvier,
       author = {{Bouvier}, J. and {Alencar}, S.~H.~P. and {Harries}, T.~J. and {Johns-Krull}, C.~M. and {Romanova}, M.~M.},
        title = "{Magnetospheric Accretion in Classical T Tauri Stars}",
    booktitle = {Protostars and Planets V},
         year = 2007,
       editor = {{Reipurth}, Bo and {Jewitt}, David and {Keil}, Klaus},
        month = jan,
        pages = {479},
          doi = {10.48550/arXiv.astro-ph/0603498},
archivePrefix = {arXiv},
       eprint = {astro-ph/0603498},
 primaryClass = {astro-ph},
       adsurl = {https://ui.adsabs.harvard.edu/abs/2007prpl.conf..479B}
}

@ARTICLE{2013Bouvier,
       author = {{Bouvier}, J. and {Grankin}, K. and {Ellerbroek}, L.~E. and {Bouy}, H. and {Barrado}, D.},
        title = "{AA Tauri's sudden and long-lasting deepening: enhanced extinction by its circumstellar disk}",
      journal = {\aap},
         year = 2013,
        month = sep,
       volume = {557},
          eid = {A77},
        pages = {A77},
          doi = {10.1051/0004-6361/201321389},
archivePrefix = {arXiv},
       eprint = {1304.1487},
 primaryClass = {astro-ph.SR},
       adsurl = {https://ui.adsabs.harvard.edu/abs/2013A&A...557A..77B}
}

@ARTICLE{2013Romanova,
       author = {{Romanova}, M.~M. and {Ustyugova}, G.~V. and {Koldoba}, A.~V. and {Lovelace}, R.~V.~E.},
        title = "{Warps, bending and density waves excited by rotating magnetized stars: results of global 3D MHD simulations}",
      journal = {\mnras},
         year = 2013,
        month = mar,
       volume = {430},
       number = {1},
        pages = {699-724},
          doi = {10.1093/mnras/sts670},
archivePrefix = {arXiv},
       eprint = {1209.1161},
 primaryClass = {astro-ph.SR},
       adsurl = {https://ui.adsabs.harvard.edu/abs/2013MNRAS.430..699R}
}

@ARTICLE{1992Attridge,
       author = {{Attridge}, Joanne M. and {Herbst}, William},
        title = "{Rotation Periods of T Tauri Stars in the Orion Nebula Cluster: A Bimodal Frequency Distribution}",
      journal = {\apjl},
         year = 1992,
        month = oct,
       volume = {398},
        pages = {L61},
          doi = {10.1086/186577},
       adsurl = {https://ui.adsabs.harvard.edu/abs/1992ApJ...398L..61A}
}

@ARTICLE{1996Choi,
       author = {{Choi}, P.~I. and {Herbst}, W.},
        title = "{Rotation Periods of Stars in the Orion Nebula Cluster: The Bimodal Distribution}",
      journal = {\aj},
         year = 1996,
        month = jan,
       volume = {111},
        pages = {283},
          doi = {10.1086/117780},
       adsurl = {https://ui.adsabs.harvard.edu/abs/1996AJ....111..283C}
}

@ARTICLE{2023Bino,
       author = {{Bino}, Gianfranco and {Basu}, Shantanu and {Dey}, Ramit and {Auddy}, Sayantan and {Muller}, Lyle and {Vorobyov}, Eduard I.},
        title = "{Predicting Stellar Mass Accretion: An Optimized Echo-State Network Approach in Time Series Modeling}",
      journal = {arXiv e-prints},
         year = 2023,
        month = feb,
          eid = {arXiv:2302.03742},
        pages = {arXiv:2302.03742},
          doi = {10.48550/arXiv.2302.03742},
archivePrefix = {arXiv},
       eprint = {2302.03742},
 primaryClass = {astro-ph.SR},
       adsurl = {https://ui.adsabs.harvard.edu/abs/2023arXiv230203742B}
}
\bibliographystyle{aasjournalv7}

\end{document}